\documentclass[10pt, conference, letterpaper]{IEEEtran}

\IEEEoverridecommandlockouts
\usepackage{cite}
\usepackage{amsmath,amssymb,amsfonts}
\usepackage{algorithmic}
\usepackage{graphicx}
\usepackage{textcomp}
\usepackage[dvipsnames,svgnames,x11names]{xcolor}
\usepackage{enumitem}
\usepackage{subfig}
\usepackage{mathtools}
\usepackage[colorinlistoftodos]{todonotes}
\usepackage{placeins}
\PassOptionsToPackage{hyphens}{url}\usepackage[bookmarks=false]{hyperref}
\usepackage{footnote}
\usepackage{footmisc}
\usepackage[final]{changes}
\usepackage{url}

\definechangesauthor[color=BrickRed]{Romain}
\definechangesauthor[color=NavyBlue]{George}
\definechangesauthor[color=Green]{Alex}
\definechangesauthor[color=RedViolet]{AlexFinal}

\newcommand{\cc}[1]{\mbox{\textbf{\cite{#1}}}}

\begin{document}

\title{Detecting Network Disruptions At Colocation Facilities
\thanks{This work has been funded by the European Research Council grant agreement no. 338402.}
}

% Add authors here

\author{\IEEEauthorblockN{Alexandros Milolidakis}
\IEEEauthorblockA{\textit{University of Crete} \\
Greece \\
alexmil@csd.uoc.gr}
\and
\IEEEauthorblockN{Romain Fontugne}
\IEEEauthorblockA{\textit{IIJ Research Lab} \\
Japan \\
romain@iij.ad.jp}
\and
\IEEEauthorblockN{Xenofontas Dimitropoulos}
\IEEEauthorblockA{\textit{University of Crete / FORTH} \\
Greece \\
fontas@ics.forth.gr}
%\and
%\and
%\and
}

\maketitle

\begin{abstract}
%This document is a model and instructions for \LaTeX.
%This and the IEEEtran.cls file define the components of your paper [title, text, heads, etc.]. *CRITICAL: Do Not Use Symbols, Special Characters, Footnotes, 
%or Math in Paper Title or Abstract.
Colocation facilities and Internet eXchange Points (IXPs) provide neutral 
places for concurrent networks to daily exchange terabytes of data traffic.
Although very reliable, these facilities are not immune to failure and may experience 
difficulties that can have significant impacts on exchanged traffic.
In this paper we devise a methodology to identify collocation facilities in
traceroute data and to monitor delay and routing patterns between facilities. 
We also present an anomaly detection technique to report abnormal traffic 
changes usually due to facilities outages.
We evaluate this method with eight months of traceroute data from the RIPE
Atlas measurement platform and manually inspect the most prominent events, that are:
an IXP outage, a DDoS attack, and a power failure in a facility.
These case studies validate the benefits of the proposed system to detect real 
world outages from traceroute data.
We also investigate the impact of anomalies at the metropolitan-level and identify
outages that span across up to eight facilities.
\end{abstract}

% Add Index keywords here

%\begin{IEEEkeywords}
%component, formatting, style, styling, insert
%\end{IEEEkeywords}

% Submitted version 31 July 2018
%\input{Submitted-31July-ver}
% Camera Ready version
% creation Date 15/6/2018

\section{Introduction}
\label{chapter:introduction}

\replaced[id=Romain]{Internet eXchange Points (IXPs) and colocation facilities}{IXPs, usually hosting access switches in colocation facilities \cc{ref:9},} have played an important role in the evolution of the Internet\cc{ref:15,ref:101}. \replaced[id=Romain]{By joining IXPs, lower tier networks are able to exchange traffic at a much lower cost than using traditional transit networks\cc{ref:13}, and at the same time experience better network performance\cc{ref:17}.}{ Lower tier networks located in different geographic locations are able to interconnect directly with each other by establishing peering links over the IXP infrastructure.}
\replaced[id=Romain]{Consequently, IXPs are now critical parts of the global Internet\cc{ref:35,ref:38,ref:26} where terabytes of traffic is exchanged daily\cc{ref:16}.
The success of IXPs is a double-edged sword as it also means that a disruptive event
at a single IXP, or one of its colocation facility, can have an impact across numerous networks.
In one example, a power outage on April 8, 2018, at a facility of one of the largest 
IXP in the world, DE-CIX Frankfurt, had broadly affected Internet connectivity 
across all Germany\cc{thousandeyes:decix-outage}.
The increasing criticality of IXPs provides a clear motivation for the development of 
techniques to detect disruptions and performance degradations at colocation facilities.
}{ These aspects led researchers, governments and intelligence agencies \cc{ref:81,ref:82} to consider IXPs as critical parts of the worldwide traffic peering \cc{ref:26}. 
On a daily basis, terabytes of traffic are exchanged through IXPs\cc{ref:16}.
However, heavy traffic anomalies, due to misconfigurations, power failures, DoS attacks, etc., can also take place between ASes over facility peering links.
The deployment of an accurate anomaly detection mechanism under facility crossings remains an important challenge.
}

\replaced[id=Romain]{This work investigates the use of data plane information to 
detect anomalies at IXP facilities. 
We leverage existing large scale traceroute results collected
by the RIPE Atlas measurement platform to model usual traffic patterns at 
colocation facilities and detect abnormal changes.
The proposed method is composed of two parts:
(i) Using IP addresses obtained in traceroutes we infer border routers connected
to IXPs and links crossing colocation facilities.
(ii) Then we identify abnormal delay and routing changes between facilities using
non-parametric statistics.
}{
This work aims to detect anomalies at IXP facilities using data plane information. 
Between May and December 2015, we analyze a huge ammount of RIPE Atlas traceroutes and perform the following steps: (i)Using the reported IP addresses we infer border routers \replaced[id=Alex]{connected to}{located in} IXPs,
then (ii)detect the facility-crossing links adjacent to the IXP hops, and finally (iii)identify delay changes and routing problems between facilities using non-parametric statistical methods \cc{ref:10}.
}

Both colocation facility detection\cc{ref:9} and abnormal delay/routing detection\cc{cing:infocom03,mahajan:sosp03,deng:icnp08,ref:10} have been separately addressed in the literature.
\replaced[id=Alex]{We study both topics jointly and in the process improve the existing methods.}
{But to our knowledge this work is the first to study both topics jointly.}
This work complements the literature with several contributions:

\begin{itemize}
    \item We design a unified method to detect facility outages, outages between 
        ASes peering at an IXP, and facility or IXP maintenances.
    \item This includes a method, called the Rule-based Constrained Facility 
        Search (RCFS), for the detection of colocation facilities in traceroute data.
    \item We also adapt existing delay and routing monitoring methods to the particular
        case of colocation facilities.
    \item The implemented system provides a unique view on inter and intra facility
        delays and packet forwarding patterns.
    \item Reported alarms are ranked by order of importance so network operators 
        can focus only on important anomalies.
\end{itemize}

\added[id=Romain]{
Our system is evaluated with eight months of traceroute data from the RIPE Atlas 
platform.
We manually check the most prominent detected outages and cross-validate our
results with information made publicly available by the community or IXP operators.
We also study the geographical span of outages and found events affecting 
multiple facilities in the same metropolitan area.}

%\todo[nolist, size=\tiny]{ rephrase  paragraph. Make clear what is the contribution?. Do I have to write about limitations?  }

The rest of this paper is organised as follows: First, we define the necessary background \textbf{(Section~\ref{chapter:background})} and our datasets \textbf{(Section~\ref{chapter:Datasets})}. Then, we describe the proposed method \textbf{(Section~\ref{chapter:Methodology})}, evaluate it \textbf{(Section~\ref{chapter: System Evaluation})} and report the existing limitations \textbf{(Section~\ref{chapter: Limitations})}. Finally, we close with the conclusions
%and the future work
\textbf{(Section~\ref{chapter:Conclusion and Future_Work})}.

%-------------------------------Section Background -------------------------------------------
\section{Background}
\label{chapter:background}

First, we introduce the peering infrastructure terminology and provide background to infrastructure outages.
%The goal of this chapter is to introduce the peering infrastructure terminology (IXPs and colocation facilities) and provide a general background on the topology discovery techniques as well as their limitations.

\subsection{Internet eXchange Point (IXP)}
\label{background:Internet exchange point}

An IXP is a network infrastructure that facilitates public peering among participant ASes \textbf{\cite{ref:13}}. IXPs usually operate at layer 2 and provide low latency and high throughput solutions to their customers.

Their infrastructure can range from a single switch with a few interconnected members to a distributed \replaced[id=Romain]{system spanning across multiple continents.}{intercontinental level.} 
\replaced[id=Romain]{Large IXPs are deployed }{Big IXPs usually install switches} in various locations (e.g., colocation centers) to provide \replaced[id=Romain]{their services to numerous networks.}{to multiple networks access to the IXP infrastructure. }
\replaced[id=Romain]{The IXP members routers are directly connected to ports of
    the IXP switches.}{
Each IXP member then brings its own router and connects one port to the IXP switch \deleted[id=Alex]{and the other to the media leading back to the AS member}.}
%\textbf{(Figure~\ref{fig:3})}.
Upon agreement of the member to the IXP terms and conditions and the successful assignment of an IXP address, the AS is ready to exchange traffic with other IXP participating networks \textbf{\cite{ref:13}}. 

%\begin{figure}%
%    \centering
%    \includegraphics[width=0.36\textwidth]{figures/Background_Figure.png}%
%    \caption{Explaining the various interconnections in colocation facilities and IXPs. Different colors belong to means different participating entities.}
%    \label{fig:3}
%\end{figure}

% \begin{figure}%
%     \centering
%     \includegraphics[width=0.40\textwidth]{figures/GR-ix.PNG}%
%     \caption{GR-IX topology in athens \textbf{\cite{ref:21}}}
%     \label{fig:4}
% \end{figure}

% \begin{figure}%
%     \centering
%     \includegraphics[width=0.40\textwidth]{figures/NL-IX.PNG}%
%     \caption{a European IXP with points of presence in various countries. Recently, a new instance has been added in Miami, Florida \textbf{\cite{ref:22}}}
%     \label{fig:5}
% \end{figure}

\subsection{Colocation Facility (Colo)}
\label{background:Colocation Facility}

%\todo[nolist, size=\tiny]{ Watch out: Do I use words from other papers  }
Colos are buildings that provide secure places for networks to bring in their equipment and interconnect.
\added[id=Alex]{Cooling, fire protection, stable power, backup generators, high bandwidth cables are only few of the services they offer to satisfy their customer needs \textbf{\cite{ref:9}}. Big IXPs usually install access switches in multiple colos in the city they operate.}
%IXPs are usually hosted by the Colos operating in the same city \textbf{(Figure~\ref{fig:6})}. 
Colo customers may then utilize the IXP infrastructure to exchange traffic with members of remote colos \textbf{\cite{ref:9}}.
\added[id=Alex]{Large companies may also operate multiple colos under the same city and connect them via cross-connect links \textbf{\cite{ref:9}}. This distributed interconnection offers better protection and stability during an outage \textbf{\cite{ref:49}}.
}

\deleted[id=Alex]{Colos provide various services to attract new customers. Cooling, fire protection, stable power, backup generators, high bandwidth cables are only few of the services they offer to satisfy their customer needs. Large companies may operate multiple colos under the same city and connect them via cross-connect links \mbox{\textbf{\cite{ref:9}}}. This distributed interconnection offers better protection and stability during an outage \mbox{\textbf{\cite{ref:49}}}.}

The peering options usually \replaced[id=Romain]{available at}{established in } Colos are:
1) Public peering, where the IXP infrastructure is utilized for the communication (\textbf{Fig.~\ref{fig:13}}A\&B), 2) Private peering, where the two peers directly exchange data either via private IXP links (tethering \textbf{\cite{ref:9}}) or by directly connecting with each other (cross-connect \textbf{\cite{ref:41}}, \textbf{Fig.~\ref{fig:13}} C,D and E) and 3) Remote peering, where a network not present in the facility remotely connects to the IXP \textbf{\cite{ref:27}}.

In this paper, we identify public peering and under certain conditions, also private peering links. The monitoring of remote peering links is not part of this work.

% Peering Infrastructure Outages
\subsection{Peering Infrastructure Outages}
\label{background:Peering Outages}

Colos and IXPs play an important role to interconnect thousands of peers around the world \textbf{\cite{ref:49}}. Power failures, human errors, attacks and natural disasters affecting these infrastructures may be \replaced[id=Romain]{critical for the Internet connectivity of thousands users}{crucial for the Internet stability for thousands of users}. Yet, only the most severe outages \replaced[id=Romain]{are publicly reported by}{become known to the public due to reports from} mailing lists (e.g., NANOG, outages mailing lists), news websites, and local operators. 
Consequently, evaluating a facility outage detection tool is challenging due to the lack of ground truth data. 

To the best of our knowledge, \textbf{\cite{ref:49}} was the first study that built an automatic tool for detecting peering infrastructure outages in near real time. The authors proposed the use of location information conveyed by BGP communities \textbf{\cite{ref:45}}, IXP and colocation websites to geolocate the source of BGP update messages. Observing multiple BGP updates in a short period, triggered their investigation module to examine if they were interconnected with the same colocation facility or IXP. Finally, they used active traceroutes as a mean to validate the disruption in the data plane.

\added[id=Romain]{
\replaced[id=Alex]{Compared to this past study our work has a more limited scope since it uses only data plane measurements. Yet for the observed facilities we expect to have better monitoring results.}
{Our work has several advantages compared to this past study.}
BGP is a control plane protocol that reveals some inter-domain connections
but based on their routing policies networks usually avoid announcing all their peering on BGP (e.g. private peerings).
Using data plane information and the large scale deployment of RIPE Atlas we strive 
to monitor more peering links and focus only on those that are actually in use. 
In addition traceroute provides RTT data that allows us to identify detrimental 
delay increases within and between facilities, for example caused by DDoS attacks
(see Section~\textbf{\ref{section DNS Root Server Attack}}).
This type of events has no impact on the control plane so it is undetected by methods using BGP data.
}

%------------------------------------ CHAPTER_3 ------------------------------------------------
\section{Datasets}
\label{chapter:Datasets}

This work aims to detect \replaced[id=Romain]{network disruptions}{traffic anomalies} at colos using traceroutes \replaced[id=Romain]{data.}{passing through the affected entity.} First we seek to identify routers located in colos and then monitor unusual \replaced[id=Romain]{routing}{traffic} and delay patterns for the facilities intra and inter links {\added[id=Alex]{(\textbf{Fig.~\ref{fig:13}})}. To achieve this, we leverage multiple datasets: 

    \textbf{RIPE Atlas \textit{built-in} and \textit{user} IPv4 Paris traceroutes measurements} from May until December 2015. 
The \textit{built-in measurements} \replaced[id=Romain]{consist in traceroutes done every 30 minutes from all Atlas probes (about 10k probes) towards all DNS root servers and a few servers operated by RIPE NCC}{ database consists of measurements from around 10,000 probes towards important destinations}\cc{ref:2}.
\deleted[id=Romain]{We focus on the IPv4 traceroutes that target instances of the DNS root servers at 30-minutes intervals.}
    \replaced[id=Romain]{In order to be closer to end-users and achieve lower latencies, numerous DNS root server instances are deployed at IXPs.}{The DNS instances tend to be close to IXPs (thus facilities) to achieve wide user coverage and  low latencies.} 
    \replaced[id=Romain]{Consequently the root DNS servers make excellent traceroute targets to monitor colos over time.}{Their distributed nature makes them excellent target candidates to monitor multiple colo paths. } \replaced[id=Romain]{The \textit{user measurements} are used for a different purpose.}{On the other hand, the \textit{user measurements} are performed on demand consuming RIPE Atlas credits.} \replaced[id=Romain]{Because they are initiated on demand and are usually lasting for a short period of time, we do not use the \textit{user measurements} }{Since, they do not have a stable pattern, we do not use this database} for anomaly detection but only to detect additional peering relationships between colo members \textbf{(Section \ref{Methodo: Facility phase})}.

    \textbf{PeeringDB \deleted[id=Romain]{database}\textbf{\cite{ref:3}}} \replaced[id=Romain]{provides diverse details about IXPs. For example, IP prefixes used for peering LANs, the facilities where IXPs are present, and ASN of member networks. This database is}{hosting independent entries} maintained by IXP and network operators. We query PeeringDB to first identify IXP addresses in the traceroute path and then extract candidate facilities. Although, our closest available snapshot is one year \replaced[id=Romain]{after the traceroute measurements}{later} (24/09/2016) we assume it is still accurate.

    \textbf{CAIDA's Internet Topology Data Kit (ITDK \cite{ref:4})}. \replaced[id=Romain]{Our facility detection algorithm seeks for IXP addresses in traceroute, but}{ IXP connected} routers do not always answer with their IXP interface \textbf{\cite{ref:11}}. We utilize ITDK's IP to alias resolution dataset to identify alias IXP interfaces in the traceroute's path. We use the aliases resolved by MIDAR \textbf{\cite{ref:5}} and iffinder \textbf{\cite{ref:6}} which yield the highest confidence with very few false positives.
Since ITDK becomes available every 6 months we use the closest snapshot produced on August 2015.

\textbf{\textit{Routeviews} prefix-to-AS map}.
\replaced[id=Alex]{
%Mapping IP addresses from traceroutes traversing inter-domain links is risky\cc{mapit:imc16}. Yet, 
PeeringDB provides AS to facility mappings. To retrieve the ASN of IP addresses found in traceroutes we use daily dumps derived from Routeviews between May and December 2015. We make such conversions from traceroutes only when it is necessary to identify the facility (Rule 3\&5 of section \ref{Methodo: Facility phase}).}
{PeeringDB provides \deleted[id=Romain]{only an} AS to facility \replaced[id=Romain]{mappings}{map}. \replaced[id=Romain]{To retrieve the ASN of IP addresses found in traceroute results}{For this reason} we use daily dumps derived from Routeviews between May and December 2015\deleted[id=Romain]{ to map network IPs to their corresponding ASN}.}

All datasets are also available for IPv6, something not taken under consideration in this study.
Networks may establish different peering decisions over IPv6 paths \textbf{\cite{ref:77,ref:78}}. Furthermore, IPv6 may be affected differently by RTT delays \textbf{\cite{ref:31}}. \deleted[id=Romain]{It is interesting to observe how they are affected during an anomaly.}
A comparison between IPv4 and IPv6 anomalies would be interesting but this is left for future work.

%------------------------------------ CHAPTER_4 ------------------------------------------------
\section{Methodology}
\label{chapter:Methodology}

% \begin{figure}%
%     \centering
%     \includegraphics[width=0.25\textwidth]{figures/Pattern.png}%
%     \caption{The traceroute and delay pattern between facility A and B}
%     \label{fig:10}
% \end{figure}

\replaced[id=Romain]{Using traceroute data we aim to detect delay and forwarding anomalies at colos.
To achieve this goal we propose to model the usual delay and forwarding patterns
observed between colos and detect deviant patterns.}
{ \replaced[id=Alex]{We seek to model the usual delay and forwarding patterns between facility members and detect the abnormal ones using traceroute data. }{This work aims at modelling usual delay and forwarding pattern between facility members and detecting abnormal ones using traceroute data.} }
%This work aims to detect delay and forwarding anomalies traffic at facilities, using data-plane information. The key idea is that if a specific delay or routing pattern is observed between facility members, during a certain time period, we expect the same behavioral pattern to also hold during the next period. To observe this pattern at small intervals, 
%\added{we use the built-in traceroutes of RIPE Atlas, performed every 30 minutes which are likely to traverse facilities. We further enhance those measurements with the anchoring traceroutes, performed every 15 minutes, to allow better visibility of colos.
%}
\replaced[id=Romain]{Specifically, the proposed system performs the following
steps every one hour:}{The proposed system performs the following steps each hour:} 
1)~It \replaced[id=Alex]{examines the \textit{built-in} measurements to detect IXP connected routers \textbf{(Section \ref{Methodo: Analysis phase})} then 
2)~identifies the colos involved in the peering communication \textbf{(Section \ref{Methodo: Facility phase} \& \ref{Methodo: Cleaning phase}}) and finally 3)~computes }
{It scans the built-in measurements to extract possible colo information \textbf{(Section \ref{Methodo: Analysis phase})} then ii)detects and extracts the pattern between colos \textbf{(Section \ref{Methodo: Facility phase} \& \ref{Methodo: Cleaning phase}}) and finally iii)proceeds with
}
delay and forwarding patterns and detects anomalies \textbf{(Section \ref{Methodo: Anomaly phase})}.

\begin{table}[]
\centering
\caption{Notations used in the methodology section.}
\label{table1}
\begin{tabular}{|l|l|}
\hline
Notation                 			& Meaning                                            \\ \hline
%(IPX - ... - IPY)          			& Sequence of IP hops in the traceroute path         \\ \hline
Alias(X)                 			& The alias interfaces of IPX (Includes IPX) \\ \hline
AS(X)                    			& The ASN of IPX                                     \\ \hline
F(X)= \{$F_A$ , ... , $F_Z$\} 	& Facilities where the network X is present          \\ \hline
\end{tabular}
\end{table}

% Analysis phase
\subsection{IXP Identification}
\label{Methodo: Analysis phase}

The goal of this step is to isolate IP addresses related to colos. Since IXPs are usually located at colos and peering LANs are easily identifiable in traceroutes, we use IXP addresses to find traceroutes traversing colos. \deleted[id=Romain]{Naturally, colos exist which don't host an IXP. \deleted[id=Alex]{Identifying addresses related to those may be possible \mbox{\textbf{\cite{ref:9}}} however,} we ignore them for accuracy reasons.
}
%Our goal in this phase is to isolate IP addresses that contain hints about the colos where the IP interface(s) are located. IXPs usually install access switches in colos and assign an address from their subnet to connected members.
%\added{Although colos exist which don't host an IXP, we consider mandatory to observe one for accuracy reasons. 
%}
%Big ISPs may covertly peer on various locations around the world. If we know that they connect to an IXP in the same location we can safely infer the facility from the IXP information.
% \begin{figure}%
%     \centering
%     \includegraphics[width=0.48\textwidth]{figures/Step1.png}%
%     \caption{Analysis phase example}
%     \label{fig:11}
% \end{figure}

In the beginning, we parse the \textit{built-in} traceroutes and extract the IP path. 
%As those measurements have been configured to run using Paris traceroute they do not suffer from the most common load balancing problems \textbf{\cite{ref:80}}. 
We sanitize it by removing hops with errors or invalid IPs (i.e. *). For the remaining clean path we query PeeringDB \replaced[id=Romain]{to check if any of the observed IP addresses belong to IXP peering LANs.}{searching if any IP belongs in the subnet of an IXP.} If such an IP is found \replaced[id=Alex]{we extract the IXP and the previous IP hop (e.g., those of IPA\&B in \textbf{Fig.~\ref{fig:14}})}{a triplet \textit{(IPA-IPIXP-IPB)} is extracted and stored in a form of pairs \textit{(IPA-IPIXP)} and \textit{(IPIXP-IPB)}}. Since an IXP appears in the path, we conclude that the \replaced[id=Romain]{traceroute traversed a public peering link.}{networks use public peering.}

Private peering is harder to identify because no IXP address appears in the path (like in \textbf{Fig.~\ref{fig:13}C,D,E}). Instead we check if alias IPs are used that are assigned to routers with IXP interfaces. \replaced[id=Alex]{Upon finding two sequential ones we extract the corresponding hops (e.g., those of IPA\&C in \textbf{Fig.~\ref{fig:13}C}).}{ If we find one we extract the two surrounding IP addresses, for example in \textbf{Fig.~\ref{fig:13}C} two pairs are extracted, \textit{(IPA-IPC)} and \textit{(IPC-IPD)}, one for each IXP connected router.}

\replaced[id=Alex]{At the end of this step, we obtain IP hops related to colos. In the next step we use the IP of the first and the IP of the second hop to detect the corresponding near-end and far-end colo. We call them near and far-end with respect to their order in the traceroute path.}{At the end of this step, we obtain IP pairs related to colos and their corresponding RTTs. In the next step we identify the actual location of each extracted IP.}

%\added{At the end of this step, we store the extracted hop pairs and their RTT delays into daily datasets. In the next phase we will use those pairs to identify the location of the peering interconnections.}

% Facility detection phase
\subsection{Facility Detection Phase (RCFS)}
\label{Methodo: Facility phase}

To identify the colos, \replaced[id=Romain]{we improve}{we modify} the constrained facility search (CFS) method\deleted{proposed in}\cc{ref:9}\replaced[id=Romain]{ to better exploit the information provided by PeeringDB.}{to prioritize specific entries of PeeringDB.}
\added[id=Alex]{The original algorithm combined IXP information from multiple sources but in doing so it ignored useful mappings between the IXP address and the AS of the customer (refer to rule 2). 
%Furthermore, since PeeringDB is the only IXP database we use and since we know that some entries are incomplete
}
%we make the following assumption: \textit{\replaced[id=Romain]{In PeeringDB the IXP to facility mappings are more reliable than the AS to facility mappings.}{IXP related data are more accurate than the network (AS) hosted ones in PeeringDB.}}

In our experiments with PeeringDB we observed that the IXP to facility mappings are more reliable than the AS to facility mappings.
%In this work, we make the following assumption: \textit{\replaced[id=Romain]{In PeeringDB the IXP to facility mappings are more reliable than the AS to facility mappings.}{IXP related data are more accurate than the network (AS) hosted ones in PeeringDB.}}
\added[id=Romain]{This is the case mainly because }IXPs \replaced[id=Romain]{tend to carefully maintain their PeeringDB entries}{utilize PeeringDB as a front-end of their costumers list} to attract new customers \textbf{\cite{ref:13}}. 
However, AS entries may be outdated or contain limited information, e.g. for security concerns. 
Based on these observations, we propose a rule (RCFS) model which strategically constrains \added[id=Alex]{the facility search for the specific case of PeeringDB}.

\replaced{Following the example of \textbf{Fig.~\ref{fig:14}} for each extracted hop we first consider the IXPs in the alias list (Rule 1\&2) then the remaining AS information of this list (Rule 3) and finally, if we could not identify the colo, the IXP and AS information of the next hops (Rule 4\&5). If possible, we avoid IP to AS conversions for border routers as they are prone to errors.
%We also avoid next hop IPs before Rule 4 because of traceroute problems, e.g. routers answering from another interface \textbf{\cite{ref:86}} or load balancing not mitigated by Paris traceroute \textbf{\cite{ref:80}}.
}
{For the near-end pairs of \textbf{Fig.~\ref{fig:14}} we first consider the IXP information of \textit{Alias(IPA)} (Rule 1\&2) then the AS information of \textit{AS(Alias(IPA))} (Rule 3) and finally, if we could not identify the colo, the IXP and AS information of the second IPs in the pairs (Rule 4\&5).
We avoid using next hop IPs before Rule 4 and 5 because they may involve traceroute problems, e.g. routers answering from another interface \textbf{\cite{ref:86}} or load balancing not be mitigated by Paris traceroute \textbf{\cite{ref:80}}.}

%\deleted[id=Romain]{
    %% We never mention this Figure
%\begin{figure}%
    %\centering
    %\includegraphics[width=0.27\textwidth, angle =90]{figures/methodo_phases.png}%
    %\caption{Facility \& Anomaly Detection phases}
    %\label{fig:12}
%\end{figure}
%}

\begin{figure}%
    \centering
    \includegraphics[width=0.50\textwidth]{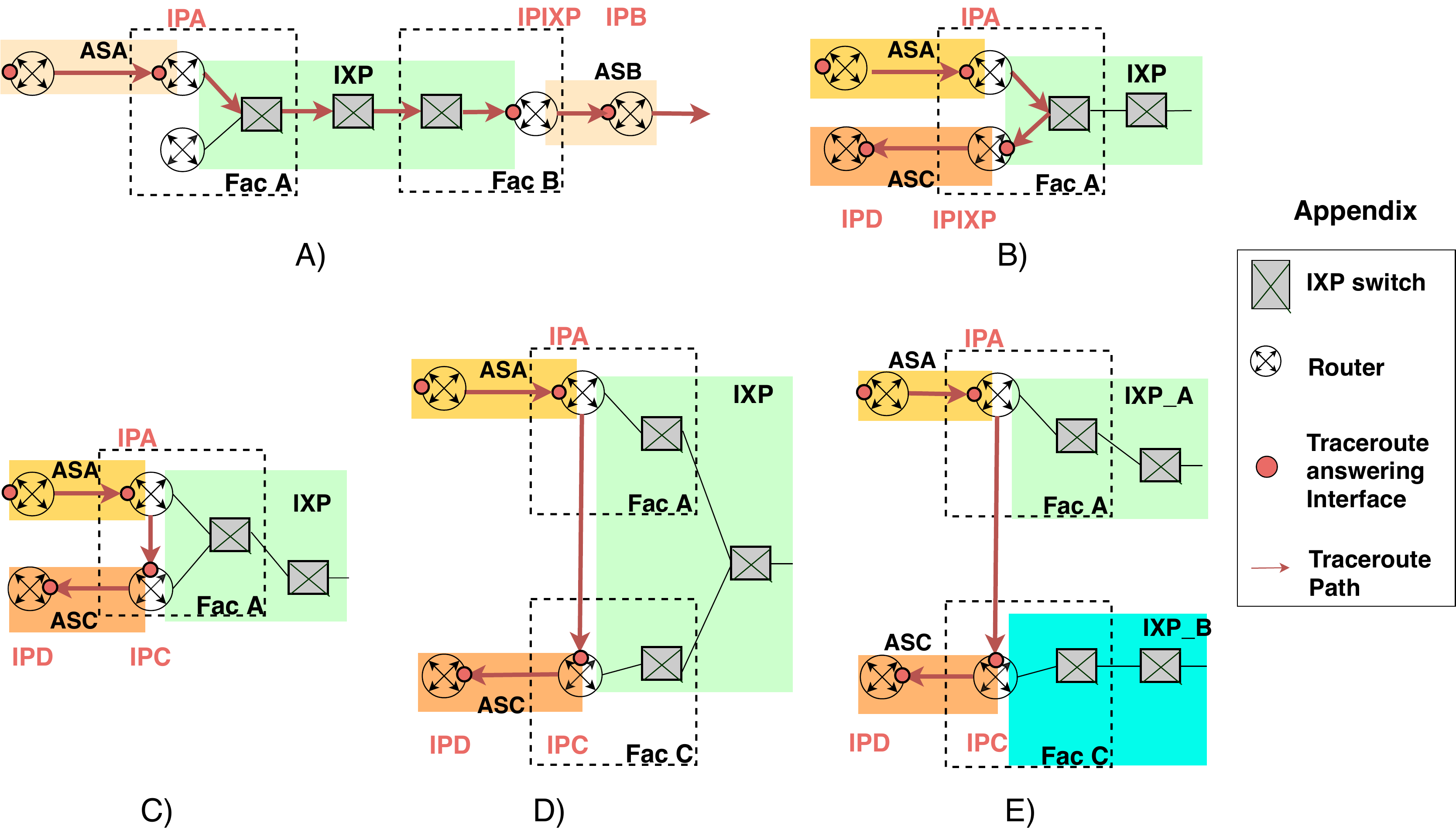}%
    \caption{Connectivity links between facilities. Public peering inter (A) and intra link (B), private peering intra (C) and inter links (D,E).}
    \label{fig:13}
\end{figure}

\begin{figure}%
    \centering
    \includegraphics[width=0.5\textwidth]{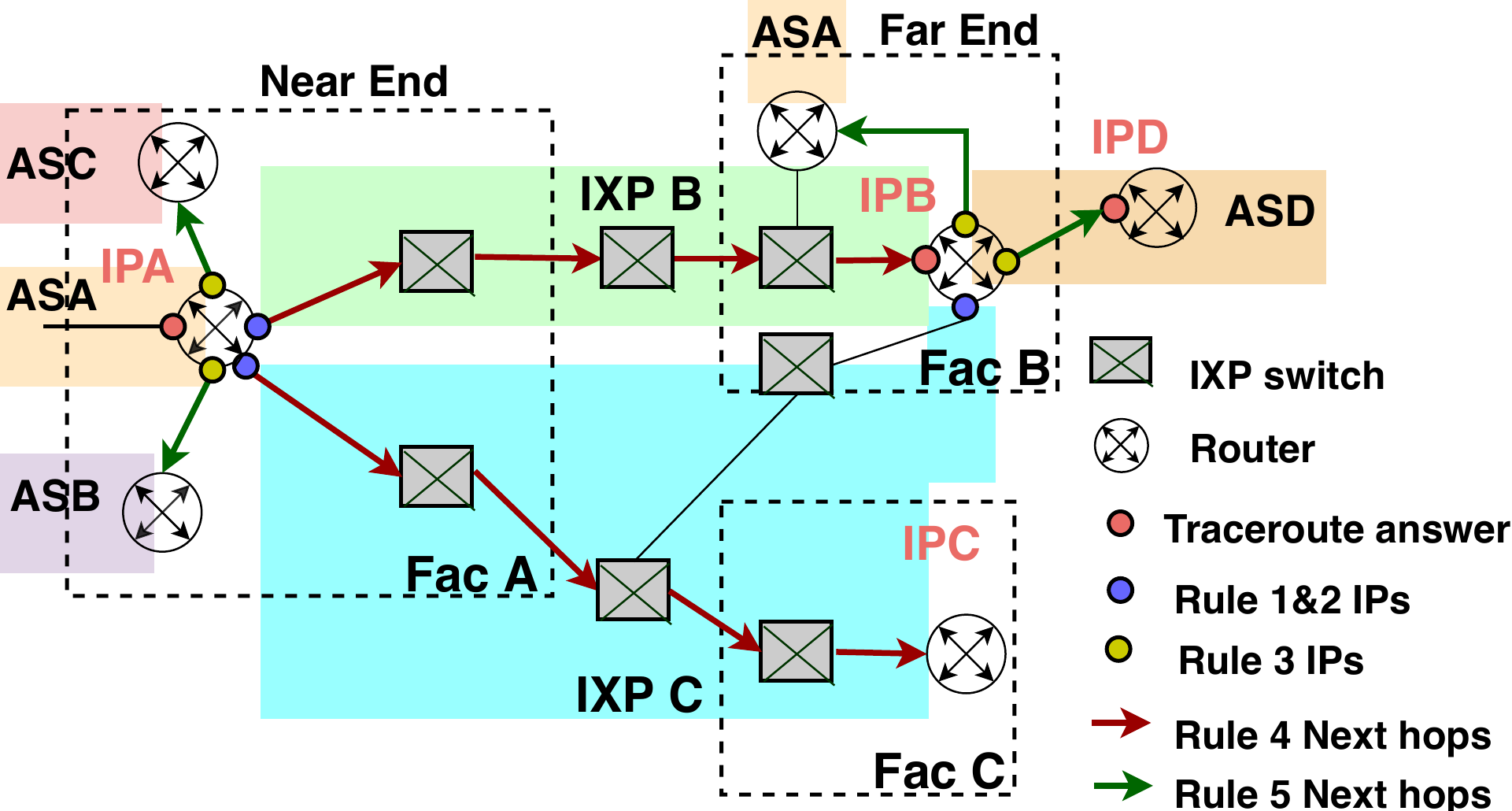}%
    \caption{Rule example of a path with extracted hops those of \textit{IPA \& IPB}. \textit{IPB \& IPC} are IXP addresses.}
    \label{fig:14}
\end{figure}

\added[id=Alex]{The detection of the near-end colo is thoroughly described below using IPA. The same method can be applied for the detection of the far-end using IPB. Each rule receives as input the candidate colos of the previous rule then constrains them and forwards them to the next rule. The algorithm stops when either a single candidate remains or none due to conflicts between rules. Note that rules can be skipped. For example, if there is no alias for an IP then R3 output is R3 input.}
\\
\textit{\textbf{Rule 0: User yielded information.}} \replaced[id=Romain]{We}{Before beginning the analysis, we} allow the user to specify IP interfaces that \replaced[id=Romain]{are known as belonging}{he already knows that they belong} to a colo router. If such an IP is found then we conclude that the traceroute traversed the specified colo. \added[id=Alex]{We use this rule to map a few known IPs close to the DNS Root servers \textbf{\cite{ref:10}} that were missing from PeeringDB.}
\\
\textit{\textbf{Rule 1: Facilities of the IXP.}} 
%\textbf{Rule 1: Constraining using the IXP.} 
We begin by looking for IXP addresses in the \textit{Alias(IPA)} list. If such addresses are found we fetch from PeeringDB all the colos of the identified IXP. Notice that a router may be connected to multiple IXPs. In \textbf{Fig.~\ref{fig:14}}, the router of IPA is connected to IXPA$\&$C. This allows to further constrain the candidate colos:
$R1=F(IXPC) \cap F(IXPB)=\{F_A, F_B, F_C\} \cap \{F_A, F_B\}=\{F_A, F_B\}$.
Since we obtained several colos, we proceed to Rule 2. 
\\
\textit{\textbf{Rule 2: Facilities of the IXP address.}}
%\textbf{Rule 2: Constraining facilities through the IXP address.} 
IXP operators report on PeeringDB the addresses they assign to their customer networks. 
If in Rule 1 we observe such an IXP address, we retrieve the customer's ASN from PeeringDB's IXP page. Then the colos where the customer is present.
%of the customer. Upon successful discovery, we again query PeeringDB but this time the facilities where the AS is present.} 
We intersect these results with the ones of Rule 1.\deleted[id=Alex]{If the intersection is empty we stop and return an error that signals conflicting informations.} 
In our example, the router owner is $ASA$
%is the router's owner to which the IXP address has been assigned 
thus $R2=R1 \cap F( ASA )=\{ F_A , F_B \}$. \added[id=Alex]{This rule was not useful for the near-end. It is important though for the far-end as it reveals the AS of the far-end connected peer which is not always visible in the traceroute}.
\\
\textit{\textbf{Rule 3: Facilities of the alias ASNs.}} 
%\textbf{Rule 3: Constraining facilities through alias ASNs.} 
\replaced[id=Alex]{When the IXP data are not sufficient we instead focus on networks peering inside the colo. First, using Routeviews we convert each non-IXP address of the \replaced[id=Alex]{\textit{Alias}}{\textit{Alias(IPA)}} list to the corresponding ASN. Then}{When neither an IXP is detected nor a single colo has been identified we change our focus from the IXP to the AS entries in PeeringDB. For each non-IXP address in the \textit{Alias(IPA)} list we obtain the corresponding ASNs from the Routeviews data and then} for each AS we fetch the candidate colos from the ASN pages in PeeringDB.
%First, we remove the IXP addresses from the \textit{Alias(IPA)} list and for
%the rest we use Routeviews to map non-IXP addresses IPs to their ASes.
%Afterwards, we query PeeringDB to find the facilities where these ASes are located. 
\replaced[id=AlexFinal]{The final results of this rule are the colos where all the ASes and the IXP (if any from R1/R2) are present.}{The candidate colos should be the ones of the previous rule, where all the alias ASes are present.} 
In our example, the router with IPA is connected to \replaced[id=Romain]{routers of}{the router of} $ASB$ and $ASC$. 
\replaced[id=Alex]{If ASB used addresses of its domain to establish the peering interconnection it is visible in the alias list thus, $R3=R2 \cap F(ASB)=\{ F_A \}$.}
{If the IP addresses used for these peering links belong to $ASB$ and $ASC$ then we can constrain the facility search using the facilities corresponding to $ASB$ and $ASC$.
Assume that this happens for $ASB$, then $R3=R2 \cap F(ASB)=\{ F_A \}$.
}

\deleted[id=Romain]{We only use networks that we can identify in this rule. We ignore IP addresses missing from Routeviews and ASNs which neither contain an entry nor report their colos in PeeringDB.}
\replaced[id=AlexFinal]{Sometimes, incomplete AS/IXP entries may return no candidate colos. We ignore those. If all R1, R2, R3 are skipped we stop the facility identification here.}{Rule 3 is a terminal rule; if no information has been revealed until this rule, we stop the facility detection. For accuracy reasons we require to have obtained at least some candidate colos by Rule 3. Rule 4\&5 focus on next hops and may be contaminated by traceroute errors.}
\added[id=Alex]{The following rules focus on the next hops. Although useful we avoid using them sooner because of traceroute problems, e.g. routers answering from another interface \textbf{\cite{ref:86}} or load balancing not be mitigated by Paris traceroute \textbf{\cite{ref:80}}.}
\\
\textit{\textbf{Rule 4: Facilities of next hop (IXP).}}
%\textbf{Rule 4: Constraining facilities through the next IXP hop.}
Midar's alias resolution has a small false positive rate yet false negatives are possible. This means that IXP addresses may be missing from the alias list in Rule 1.
%\replaced[id=Romain]{In previous rules we may miss the IXP used for the peering link if the \textit{Alias(IPA)} addresses do not map to the IXP. }{Although MIDAR has very few false positives, IP interfaces may still remain undetected \cc{ref:5}.} 
Assuming that routers answer with their inbound interface, the IXP of the near-end is surely observed after crossing the IXP link (IPB$\&$C in \textbf{Fig.~\ref{fig:14}}).
\replaced[id=Alex]{Rule~4 takes advantage of this observation to constrain with colos of undetected IXPs that}{identify IXPs that remained undetected in the alias list but} appeared in the next hop. We consider as next hops all those that appeared in the \textit{built-in} and \textit{user} defined measurements during the same day.\\
%\added{When we try to detect the facility of the near-end peer the IXP used for the peering interconnecton is usually observed in the next hop (e.g., IPB \& IPC are IXP addresses in fig~\ref{fig:14}). Since, this IXP is present in the near-end facility we can further constrain the facilities of rule 3 with the IXP facilities of the next hop. Rule 4 takes advantage of this observation to identify IXPs that were not observed in the \textit{Alias(IPA)} list. Although MIDAR has very few false positives, IP interfaces may still remain undetected \textbf{\cite{ref:5}}. We consider as next hops of IPA all those that appeared in the build-in and user defined measurements during the same day. Note that we use the user measurements only to discover additional peering links between colo members. We don't monitor those links for the anomaly detection.
%}
\textit{\textbf{Rule 5: Facilities of next hop (AS).}}
%\textit{\textbf{Rule 5: Constraining facilities through the next AS hops.}}
As a last resort, \added[id=Alex]{following the idea of Rule 4} we pick all the alias\replaced[id=Alex]{(es)}{( interfaces} of IPA and for each IP we resolve the next hop AS(es).
Our goal is to reveal additional peerings between $ASA$ and networks that were not observed in Rule~3. 
\replaced[id=Romain]{Then we}{If we find such a relationship we can then} utilize the colos of \replaced[id=Romain]{these ASes to constrain our facility search.}{the other peer to constrain the ones remaining.}
However, \replaced[id=Romain]{for colos that are geographically close, cross-connect links may cause wrong inferences}{problems may arise when multiple colos exist close to each other. Cross-connect links may result the other peer to be member of another colo} (e.g., \textbf{Fig.~\ref{fig:13}D\&E}).
In order to mitigate this problem, we independently intersect each new \replaced[id=Alex]{peer's}{peer AS} colos with the results of Rule~4. 
\replaced[id=Alex]{Among the independent intersections we pick as candidate the colo which appeared in the majority if it accounts at least 75$\%$ of those intersections.}
{Then we count the number of new peer ASes per colo, and consider a colo as successfully identified if it account for at least 75\% of these ASes.}

%For example, assume that $R4=\{F_A , F_B, F_C\}$ and we observe four new ASes, $AS1,AS2,AS3,AS4$, with candidate colos: 
%$F(AS1)=\{ F_A , F_B , F_D \}$, $F(AS2)=\{ F_A , F_B , F_C \}$, $F(AS3)=\{ F_B , F_E \}$, $F(AS4)=\{ F_C \}$.

%Then the common facility sets with R4 are:\\
%$F(AS1) \cap R4 = \{ F_A , F_B \}$,
%$F(AS2) \cap R4 = \{ F_A , F_B \}$
%\\
%$F(AS3) \cap R4 = \{ F_B \}$, and
%$F(AS4) \cap R4 = \{ F_C \}$.

%In this case the most common colo among those sets is $F_B$  with a score of $3/4 = 75\%$. Since this score is equal to the threshold we conclude that the IP is located in this colo. 

\added[id=Romain]{The 75\% threshold is empirically found with our traceroute dataset.}
\replaced[id=Alex]{Compared to the 100\% threshold which discards 32\% of the IPs the 75\% discards 26.7\% as depicted in \textbf{Fig.~\ref{fig:32x}} (left). This allows identification of a few additional colos, 5.3\% in total.}
{Between May and December 2015 the 75\% threshold allows identification of a few additional colos, 5.3\% in total as depicted in \textbf{Fig.~\ref{fig:32x}} \added[id=AlexFinal]{(left)}.
}
Although, there is the possibility of some incorrect inferences, we consider this threshold beneficial for identifying routers used to establish multiple peering relationships.

%\begin{figure}%
%    \centering
%    \includegraphics[width=0.33\textwidth]{figures/Failed_to_Total.png}%
%    \caption{ Percentage of failures based on a sliding threshold value. Setting the threshold to 50$\%$ means that every pair is converged. Setting it to 99$\%$ means that only perfect score pairs are converged (32$\%$ of the pairs are lost)}
%    \label{fig:32}
%\end{figure}
%
%\begin{figure}%
%    \centering
%    \includegraphics[width=0.36\textwidth]{figures/Failed_to_Total_Data_cleaning_Phase.png}%
%    \caption{Percentage of dropped IPs based on the threshold value of the data cleaning phase. Setting the threshold 99$\%$ means that only IPs which always match in the same facility will be accepted.}
%    \label{fig:33}
%\end{figure}

\begin{figure}%
    \centering
    \includegraphics[width=0.45\textwidth]{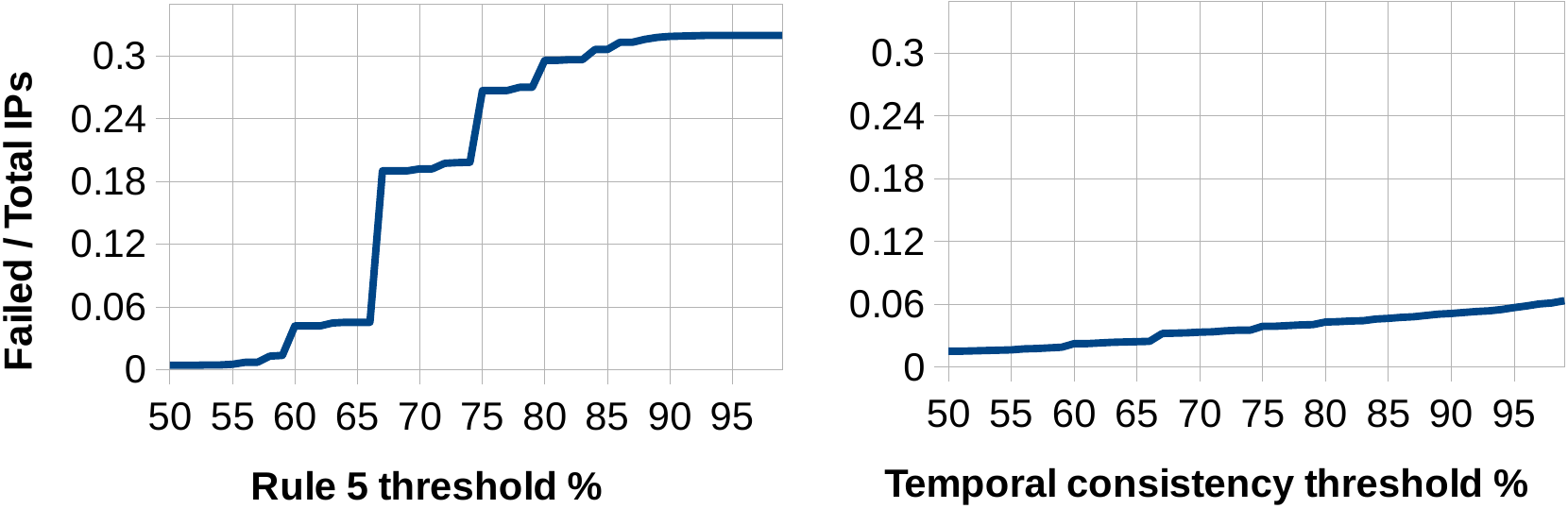}%
    \caption{Dropped IPs based on the sliding threshold. With a 50$\%$ threshold every IP is accepted meanwhile with a 100$\%$ threshold only consistent IPs are accepted; 32$\%$ (left) and 6.3$\%$ (right) are lost.}
    \label{fig:32x}
\end{figure}

\replaced[id=Alex]{At the end of this step, we temporarily store the near-end and far-end identifications. In the next step we make the final colo decision. Note that for the case of intra-colo links the near-end and far-end colos are the same (like in \textbf{Fig.~\ref{fig:13}B\&C}).}
{Upon successful identification of the near-end colo, we repeat the algorithm using as input the far-end IP pairs, that is the pairs having IPB as first hop in our example (Fig.~\ref{fig:14}). If this step is successful, the pairs are transformed to their near-end and far-end colos.
In the case of intra-colo links the far-end and near-end colo are the same like in \textbf{Fig.~\ref{fig:13}B\&C}.}
%The final choice, whether this decision is correct or not, will be taken in the data cleaning phase.  

% Facility detection phase
\subsection{Temporal Consistency}
\label{Methodo: Cleaning phase}
\replaced[id=Romain]{Some of our datasets are daily updated, hence outlier values may temporarily appear and punctually impact the facility identification results.
To address this issue we check the stability of the IP \replaced{to facility mappings}{pairs} across time and clean aberrant results. 
\textbf{Fig.~\ref{fig:32x}} (right) depicts that only a few IPs (6.3$\%$) are unstable.}
{The goal of this step is to check the stability of the IP pairs across time and clean aberrant results. 
Although all the pairs that have either IPA or one of its aliases as the first hop are consistent within the same day this is not guaranteed for individual dates. 
Each day we load a different next hop database and Routeviews dataset. 
For ths reason, we need to parse the output of the previous step an additional time. 
For each of the pairs above we pick as correct the colo where they matched in the majority of dates, above a 87$\%$ threshold. 
Pairs that failed to pass threshold are removed from the detected dataset.} 
\deleted{\textbf{Fig.~\ref{fig:32x}} (right) depicts that only a few IPs (6.3$\%$) are unstable. 
The 87$\%$ threshold allows to map 1.5$\%$ of those to their colo. More conservative studies can set the thresholds to higher values.
}
\deleted[id=Alex]{The goal of this step is to check the stability of the previous results across time and clean aberrant results. Each day we load a different next hop database and Routeviews dataset. This may result in a slightly different information causing some IP addresses either to fail the detection or to match on a different colo.
}

%The goal of the data cleaning phase is to analyze the facilities inferred in the previous phase and issue the final decision of whether the pair is safe to use or discard it. When analyzing the pairs, each day we load a different next hop database and Routeviews dataset. This may result in a slightly different hop and AS information causing some IP addresses either to fail the facility detection or to match on a different one.
%, usually situated in the same city, close to the original 

\deleted[id=Alex]{We parse the output of the previous step an additional time. For each pair $\{IPA, *\}$, we pick as correct the colo where IPA was matched the majority of times. We use a 87$\%$ threshold under which we remove the pair from the detected dataset.
}
\deleted[id=Alex]{\textbf{Fig.~\ref{fig:32x}} depicts that only a few IPs (6.3$\%$) are unstable. The 87$\%$ threshold allows to map 1.5$\%$ of those to their colo. More conservative studies can set the thresholds to higher values.
}

% Anomaly Detection Phase 
\subsection{Anomaly Detection}
\label{Methodo: Anomaly phase}

%If we represent the detected facilities as vertices we draw direct edges between them to illustrate the near and far-end facilities. 
%Those links are critical; 
Links between colos are critical. \replaced[id=Romain]{Disruptions on these links}{an anomaly there} may cause connectivity problems to thousand \added[id=Alex]{Internet} users. 
\replaced[id=Alex]{We adjust the techniques of \textbf{\cite{ref:10}} and build a simple tool to detect abnormal patterns for the specific case of colos.}{We build a simple tool to monitor delay and forwarding patterns and report abnormal patterns.}
First, \replaced[id=Alex]{we compute}{it computes} the forwarding pattern of each colo (\textbf{Section \ref{section: Packet Forwarding Model}}) then, the delay of each near-end link towards the far-end (\textbf{Section \ref{section: Delay change detection}}) and, lastly \replaced[id=Alex]{we compare}{it compares} those two patterns to computed references to detect anomalies \added[id=Alex]{(\textbf{Section \ref{Section: Reference computation}})}.

%Traceroutes of course, do not reveal the full truth about how the user traffic is affected however, they are one of the few representative tools an operator has to identify network anomalies.

% \begin{figure}[!h]%
%     \centering
%     \includegraphics[width=0.25\textwidth]{figures/AnomalyDetectionModule.png}%
%     \caption{ A sum up of the anomaly detection phase }
%     \label{fig:19}
% \end{figure}

% Anomaly subsections
\subsection{Forwarding Model}
\label{section: Packet Forwarding Model}

% \begin{figure}[!h]%
%     \centering
%     \includegraphics[width=0.15\textwidth]{figures/fwd_Anomaly_Romain.png}%
%     \caption{  Forwarding pattern of a Router's Interface R. B and C are the next hops observed in the traceroutes. Z represents packets to unresponsive interfaces. }
%     \label{fig:20}
% \end{figure}

%\added{Under normal conditions, we observe a usual pattern of packets exchanged between colo networks. During an anomaly though, packet loses and BGP changes can cause this pattern to deviate from normal.}
\replaced[id=Alex]{We collect traceroutes for each Atlas probe, extract the colos, and count the number of times links between colos are traversed. From all different probes, we aggregate those counters to produce the forwarding pattern of each near-end colo. %Probes monitor in 30 minutes intervals so this pattern should be stable. 
We compare it to a computed reference that represents usual patterns (see \textbf{Section~\ref{Section: Reference computation}}).}
{We propose a forwarding model to monitor the number of packets passing from a near-end facility to its neighboring facilities. We call this the forwarding pattern of the near-end colos. These patterns are compared to a computed reference that represents usual patterns (see \textbf{Section~\ref{Section: Reference computation}}).}

\textbf{Figure~\ref{fig:21}} illustrates an example of 
this pattern for the near-end (A) towards each far-end (B,C,D,E). The usual forwarding pattern computed at hour t-1 is $F^A = [100, 250, 80, 70]$. In the next hour, we notice an unusual decrease in the packets towards B and C accompanied by a similar increase towards facility D.

\replaced[id=Romain]{To detect anomalous patterns, we test for homogeneity with the chi-squared test and the following null hypothesis:}{We formulate a Null hypothesis and use the chi-squared to test for homogeneity.}\\
\textit{\textbf{Null hypothesis}: The observed data for an hour is consistent with the normal reference computed from the previous hours.}\\
\textit{\textbf{Alternative}: The observed data is not consistent with the normal reference.}

\begin{figure}%
    \centering
    \subfloat[Time t-1]{{\includegraphics[width=0.11\textwidth]{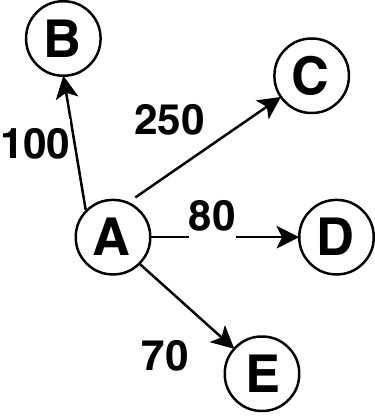}} }%
    \qquad
    \subfloat[Time t]{{\includegraphics[width=0.11\textwidth]{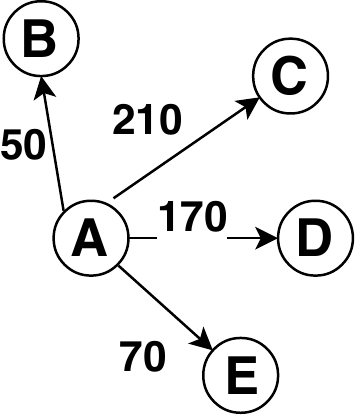}} }%
    \caption{Usual (a) and anomalous (b) forwarding patterns for colo A towards the far-end colos B, C, D and E.}%
    \label{fig:21}%
\end{figure}

%The Chi square statistic is a simple lightweight metric that we use to detect forwarding anomalies. When traffic enters A it will be forwarded either to B,C,D or E. Defining this forwarding choice as a categorical variable, one-way (goodness-of-fit) chi-square test can be used. Normally, this type of test involves the comparison with a theoretical expected distribution [\textbf{~\ref{fig:12}}]. 
%\added{In our case though, new BGP policies can force this pattern to change. Instead, we use the reference pattern of the previous hour as the ideal pattern of our distribution (\textbf{fig ~\ref{fig:21}a}). We formulate the following hypothesis:
%}
%\\
%\textit{\textbf{Null}: The observed data of this hour are consistent with the expected distribution calculated from the previous hour.}\\
%\textit{\textbf{Alternative}: The observed data are not consistent with the expected distribution}

\deleted[id=Romain]{Our goal is to search for time periods where the Null hypothesis is rejected in favor of the alternative. }
\added[id=Alex]{Since Atlas probes traceroute the same destination every 30 minutes,}
under normal network conditions, we expect the current pattern to be consistent with the distribution of the reference. 
For the chi-squared test we set a significant level of 0.01 under which we reject the null hypothesis and report an alarm. Since, this type of test does not work properly for small expected values ($<5$), \replaced[id=Romain]{we sum counts from far-end colos with less than 5 packets into one variable.}{we group all those links under a single variable expressing the sum of packets crossing through their links.}

%Large colos with several multiple IXPs can have multiple far-end facilities.
Usually only few paths are responsible for a forwarding anomaly. Suppose $F = \{ p_i \in [1, n] \}$ is the anomalous pattern and $\overline{F} = \{ \overline{p}_i \in [1, n] \}$ the computed reference. We reuse the responsibility metric defined in \textbf{\cite{ref:10}}
%\todo[nolist, size=\tiny]{fix:same but without the pearson coeff}
to detect which path caused an anomaly:
\[ r_i = \frac{ p_i - \overline{p}_i }{ \sum_{j=1}^{n} |p_j - \overline{p}_j| }\]
The responsibility metric values range from [-1, 1]. Negative values stand for paths with an unusually low number of packets. Positives values represent an unusually high and values close to zero for normal situations.

% Anomaly subsections
\subsection{Delay Change Detection}
\label{section: Delay change detection}

\replaced[id=Romain]{
Estimating delays for intra and inter facility links is not a trivial task because of traceroute limitations, such as path asymmetry and RTT variability \textbf{\cite{vries:aims15,schwartz:infocom10,ref:10}}.
}
{
The proposed detection model monitors an estimator of the delay required for the traceroutes packets to traverse intra and inter facility links.
An accurate estimation that a link is affected by delay is not trivial due to the path asymmetry problem \added[id=Alex]{(need ref. Any idea?)} }

In \textbf{\cite{ref:10}} a technique is proposed to address these challenges when a link is observed by a sufficient number of probes with different return paths. That technique monitors the shifts in the distribution of the median differential RTT($RTT_{Diff}$) and distinguishes strong alarms.
\replaced[id=Alex]{Since, colo links are usually monitored by multiple probes from different ASes with disparate return paths, we implement that monitoring technique.}
{We implement their methodology since the facility links are usually monitored by multiple probes from different ASes.}

From \textbf{Section \ref{Methodo: Analysis phase}} we extracted the IP hops and the RTT values. In \textbf{Section \ref{Methodo: Facility phase}} we found the facilities. So for the traceroute of \textbf{Fig.~\ref{fig:14}} $RTT_{Diff}=RTT_{far end} - RTT_{near end}=RTT_{IPB} - RTT_{IPA}$. We group all $RTT_{Diff}$ values from the same near-end towards the same far-end colo and calculate the median over those to ensure that an anomaly will trigger only if the majority of RTT values between the two facilities get affected.
\added[id=Alex]{Note that multiple routers of colo A and colo B may be involved in this grouping.}
\added[id=AlexFinal]{Like in \textbf{\cite{ref:10}}, we calculate confidence intervals for both the observed and the reference $RTT_{Diff}$ to detect significant statistical changes. If those confidence intervals stop to ovelap we report an alarm like those of \textbf{Fig.~\ref{fig:36}}.}

\subsection{Reference Computation}
\label{Section: Reference computation}
\replaced[id=Romain]{The normal reference to detect anomalies is computed every hour using exponential smoothing for both the forwarding and delay patterns:}
{Each hour we compute a reference value and compare with the current observed values to detect anomalies. We do this for both the forwarding and the $RTT_{Diff}$ pattern using exponential smoothing:}
\[ m_t^{Ref} = am_t + (1-a)m_{t-1}^{Ref}, \quad F_t^{Ref} = aF_t + (1-a)F_{t-1}^{Ref}\]
%\[F_t^{Ref} = aF_t + (1-a)F_{t-1}^{Ref}\]
Where $m_t^{Ref}$ is the reference $RTT_{Diff}$ of the monitored link, and $F_t^{Ref}$ the reference forwarding pattern of the facility. Likewise, $m_{t-1}^{Ref}$ and $F_{t-1}^{Ref}$ are the reference patterns of the previous hour and, $m_t$  and $F_t$ are the current observed values. The exponential smoothing parameter $a \in (0,1)$, controls the importance of new measures as opposed to previous observed ones. In our case, we set a = 0.03 to mitigate faster the impact caused by sudden anomalous bursts.

The initial reference value $m_{0}^{Ref}$ and $F_{0}^{Ref}$ are quite important when $a$ is small. To solve the cold start problem, we calculate them over the median of the values observed during the first day of our analysis. 
\replaced[id=Romain]{We maintain a different reference for each facility and update them at each one-hour time bin.}{For each facility we maintain the reference in memory and updates them with new observations.}

%------------------------------------ CHAPTER_6 ------------------------------------------------
\section{System Evaluation}
\label{chapter: System Evaluation}

Now we evaluate our proposed rules and their assistance in identifying facilities.
%In this section, we evaluate our system implementation and explain some of our decisions made during the facility detection. We analyze each of our proposed rules and how much each assisted for the identification of the facility.
Then, we discuss the anomaly detection results and present the most significant detected disruptions. Anomaly detection systems are usually evaluated in terms of true positives and false posititives, however, in our case such validation is challenging since confidential information is needed both for the facility detection and alarm validation. 
%Operators usually consider those data private for security reasons. Some IXPs \textbf{\cite{ref:84}} may reveal IP to facility related information however, the age of our dataset makes validation a difficult task.

We verify the correctness of our anomaly detection system by checking our top reported alarms, and the public reports of major outages that took place in 2015. In our results we report anomalies caused by an IXP outage on May, a power failure in colocation facility at mid November and a DDoS attack at the end of November.

\subsection{RCFS Evaluation}
\label{section: Methodology Evaluation}

We evaluate our facility identification method, RCFS, with the \textit{built-in} measurements from May until December 2015. On average, each day we analyze 12 million traceroutes and extract
\replaced[id=Alex]{14.500 unique router interfaces potentially located in colos forming 14 million IP interconnections. This large number of interconnections  is due to the rich inter-IXP connectivity \textbf{\cite{ref:44}}.}
{14 million IP pairs corresponding to 14.500 unique router interfaces that are potentially located in colos. Recall that for each IXP address in the traceroute path, we extract two pairs (e.g. \textit{(IPA, IPIXP)} and \textit{(IPIXP, IPB)}). We use IPA of the first pair to identify the near-end facility and IPIXP of the second pair to identify the far-end. The second IP in each pair assists only in the facility discovery, i.e., in Rule 4 and 5. The difference between the number of compared pairs and interfaces is due to the rich inter-IXP connectivity~\textbf{\cite{ref:44}}.}  

\replaced[id=Romain]{To understand the contribution of each RCFS rule, we picked
a smaller dataset, from May to August, and inspected in details the identification
results per rule.
For this smaller dataset RCFS consistently identifies facilities for 4000 of the 14500 interfaces (28\%).}{
In a snapshot of 4 months, between May and August, our algorithm is consistent each day identifying 4000 (28\%) out of the 14500 interfaces. 
We picked a smaller dataset to perform this evaluation however, similar results should apply for the whole 2015 period.}
%Compared to our algorithm, 
%the original CFS methodology in \textbf{\cite{ref:9}} maps 9812 interfaces. Our goal was not to perform a traceroute and IXP campaign like in \textbf{\cite{ref:9}} since combining and constantly updating IXP and facility information is a tedious process \textbf{\cite{ref:12}}. Although, we lack completeness, we can still infer anomalies in colos.

% \begin{figure}[!h]%
%     \centering
%     \includegraphics[width=0.45\textwidth]{figures/Resultsfigure1_new.png}%
%     \caption{ Number of interfaces identified each day between May and august. }
%     \label{fig:26}
% \end{figure}

% \begin{figure}[!h]%
%     \centering
%     \includegraphics[width=0.45\textwidth]{figures/Resultsfigure2_new.png}%
%     \caption{ Percentage of interfaces identified each day between May and august }
%     \label{fig:27}
% \end{figure}

\textbf{Table~\ref{table: Eval}} and \textbf{Fig.~\ref{fig:28x}} depict the contribution of each rule for the facility identification. 
We observe that the first three rules are responsible for 39$\%$ of the detected facilities. 
For the rest 61$\%$, \replaced[id=Alex]{next hop information is required (Rule~4\&5).}
{Rule~4\&5 require information from the next hop.}
Rule~4 is responsible for the majority of identifications since, the near-end colo requires knowledge of the IXP whose IP interface usually appears in the next hop, \added[id=Alex]{e.g. like in \textbf{Fig.~\ref{fig:13}A\&B}}.
In our experiments we also used Rule~0 to map a few IPs ($<0.1\%$) 
\replaced[id=Alex]{not listed in PeeringDB but on the IXP websites.}% to their facilities.}
{Those IPs were not listed in PeeringDB but were in the IXP websites.} 

\begin{table}[!t]
\centering
\caption{Average facility converges \& failures by each rule.}
\begin{tabular}{|l|l|l|l|l|l|l|}
\hline
\textit{Average \%}   & \textit{Rule 0} & \textit{Rule 1} & \textit{Rule 2} & \textit{Rule 3} & \textit{Rule 4} & \textit{Rule 5} \\ \hline
\textit{Successes \%} & \textit{0.097}  & \textit{5.391}  & \textit{18.188} & \textit{15.289} & \textit{46.731} & \textit{14.301} \\ \hline
\textit{Failures \%}  & \textit{0}      & \textit{0.604}  & \textit{2.803}  & \textit{23.823} & \textit{19.224} & \textit{53.543} \\ \hline
\multicolumn{7}{|l|}{\textit{Unique Router Interfaces observed per day:       $\sim$14.500}}                                      \\ \hline
\end{tabular}
\label{table: Eval}
\end{table}

%\begin{figure}[!h]%
%    \centering
%    \includegraphics[width=0.40\textwidth]{figures/Resultsfigure4_new.png}%
%    \caption{ Daily percentage of uniques IP that converged by each rule.}
%    \label{fig:28}
%\end{figure}
%\begin{figure}[!h]%
%    \centering
%    \includegraphics[width=0.40\textwidth]{figures/Resultsfigure5_new.png}%
%    \caption{ Daily percentage of unique IPs that failed to converge by each rule.}
%    \label{fig:29}
%\end{figure}

\begin{figure}%
    \centering
    \includegraphics[width=0.49\textwidth]{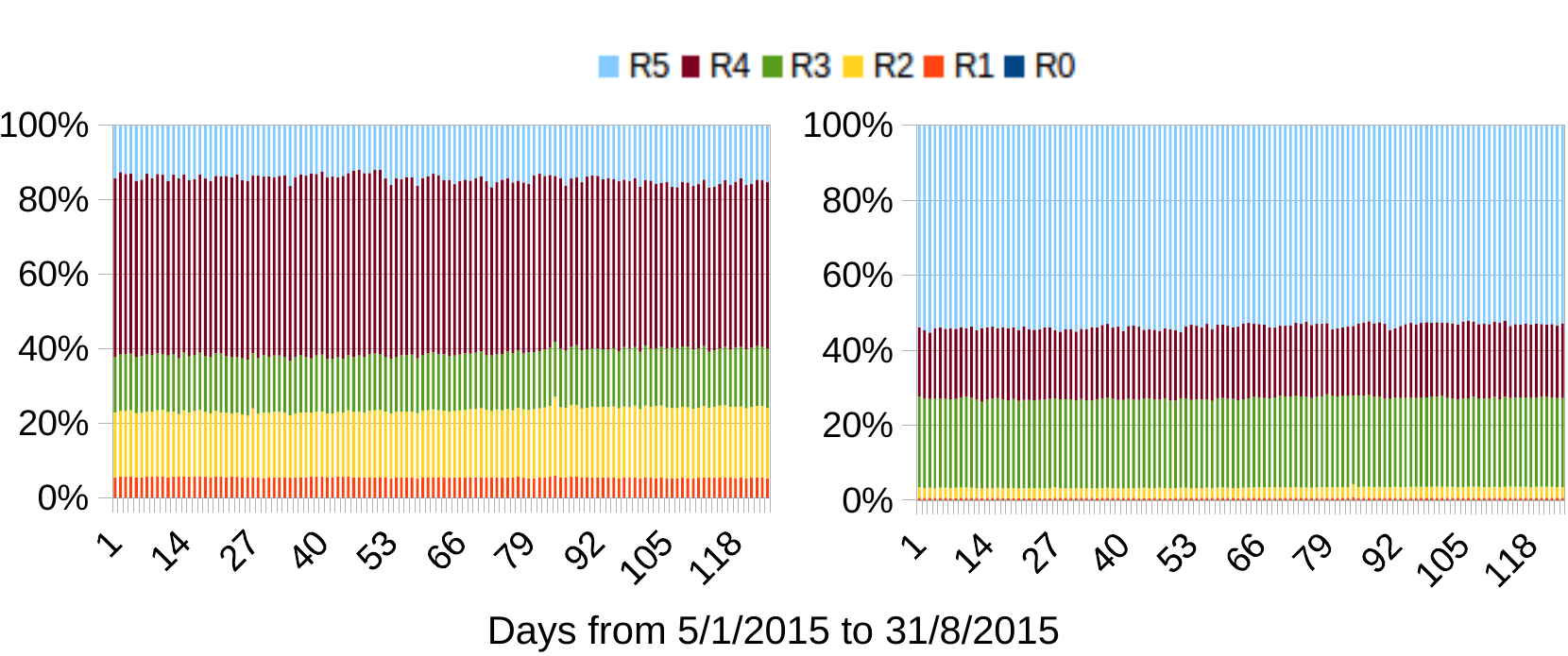} % 0.44
    \caption{Daily $\%$ of unique IPs whose facility was identified be each rule(left) or failed to do so(right).}
    \label{fig:28x}
\end{figure}

\added[id=Alex]{Similarly,}
\replaced[id=Romain]{the right-hand side plot of \mbox{\textbf{Fig.~\ref{fig:28x}}}}{Similarly, \mbox{\textbf{Fig.~\ref{fig:28x}}} right}
reveals the rules where the identification failed. RCFS fails identifying the colo when the search space becomes empty. \replaced[id=Alex]{Failures at the last rule 5 usually mean that we are unable to detect sufficient peering relationships to constrain the candidate colos.}
{a failure in Rule 5 may also mean that the pair was unable to converge due to insufficient information.}
As shown in \textbf{Fig.~\ref{fig:30x}}, this is the case for about 85\% of the IPs. For these interfaces we would require additional active traceroutes \added[id=Alex]{either from RIPE Atlas or from other sources (e.g. CAIDA Ark).} 

From the above figures, we clearly observe that our results are consistent and stable over time. This is required for the anomaly detection to safely identify pattern discrepancies. 

% \textbf{Figure~\ref{fig:27}} shows that our results are consistent proving that the facility paths chosen by the RIPE Atlas traceroutes remain stable over time. This is required for the anomaly detection module to safely identify pattern discrepancies. 

% \textbf{Figure~\ref{fig:29}} reveals the rules responsible for most of the facility detection failures. A pair to fail to converge after applying the first four rules means either that the PeeringDB data are incomplete/incorrect or that our alias information hosted is corrupted by falsely inferred next hops. A failure in rule 5, does not always mean that the pair was rejected due to false information; it might happen when the next hop ASes have been rejected due to a high threshold value. \textbf{Figure~\ref{fig:31}} analyze this case. We usually see that more than 71$\%$ of the IPs that have their facilities inferred by rule 5, have been perfectly converged perfectly without applying a threshold. 

%\begin{figure}[!h]%
%%    \centering
%%    \includegraphics[width=0.40\textwidth]{figures/Rule5ThresholdvsConvergence_new.png}%
%%    \caption{Percentage of IPs which failed to converge by rule 5 versus those that failed the threshold of this rule.}
%%    \label{fig:30}
%%\end{figure}

\begin{figure}
\centering
\begin{minipage}{0.485\columnwidth}
  \centering
  \includegraphics[width=\columnwidth]{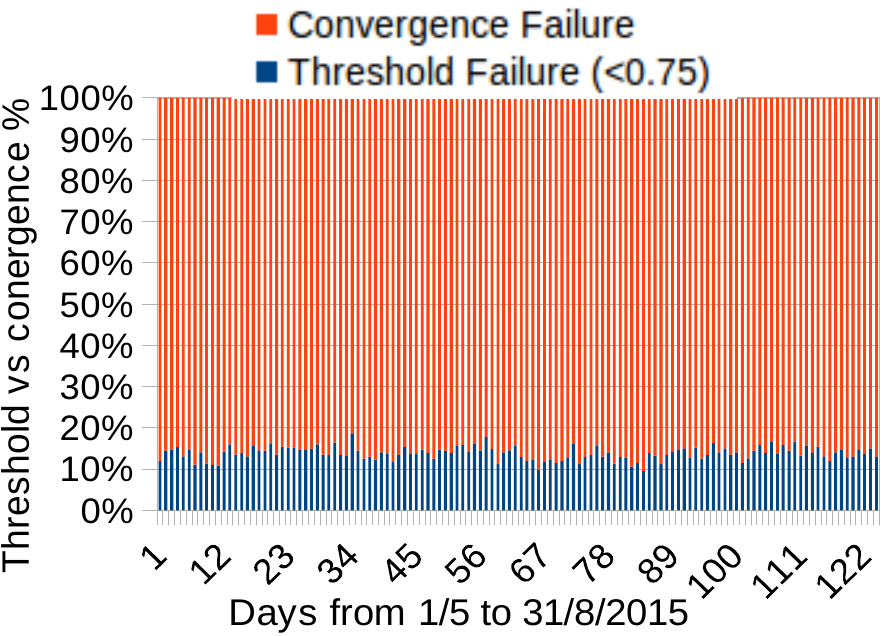}
  %\captionof{figure}{$\%$ of IPs that failed Rule 5 versus those that failed the threshold.}
  \captionof{figure}{Covergence vs threshold failure for the IPs of rule5}
  \label{fig:30x}
\end{minipage}%
\hfill
\begin{minipage}{0.50\columnwidth}
  \centering
  \includegraphics[width=\columnwidth]{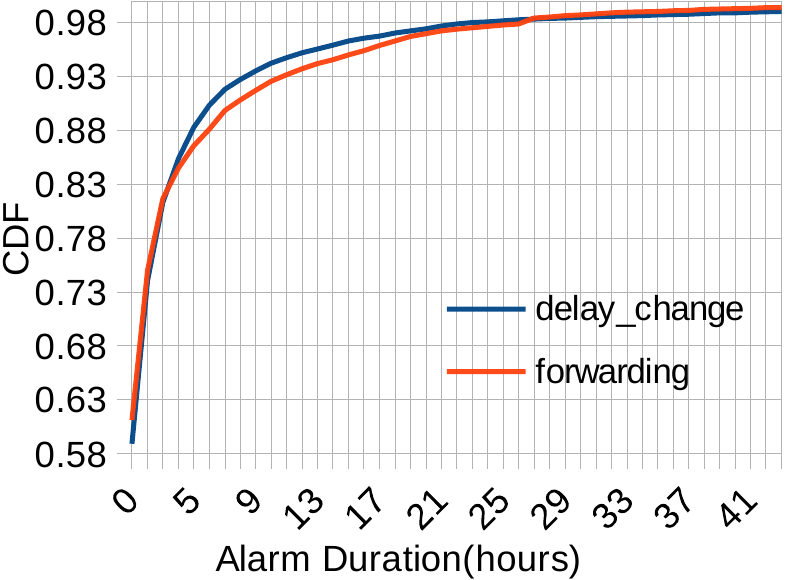}
  %\captionof{figure}{The Duration of alarms reported by the detection modules.}
  \captionof{figure}{The Duration of alarms of the detection modules.}
  \label{fig:21xx}
\end{minipage}
\end{figure}

\subsection{Anomaly Detection Evaluation}

\deleted[id=Alex]{From the 4000 daily interfaces we observe in total 264 facilities where both the near-end and the far-end facility were identified. Identification of the far-end is challenging since the far-end router typically replies with the IXP interface. The AS connected one usually does not appear in the path and the next hop interface might belong to a different AS (\textbf{section \ref{section: Fac Alg Limitations}}).}

\replaced[id=Alex]{From the 4000 daily interfaces between May and December 2015 we monitor links between 264 facilities. From those colos, 156 were reported as anomalous at least once.}
{Between May and December 2015, we observe 156 out of the 264 colos to be reported at least once.}
%out of the 264 we detect 156 facilities to be affected at least once.
In total {we found that }the observed patterns deviate from the computed references 13135 times for the forwarding analysis and 19850 times for the delay analysis.
61$\%$ of the forwarding alarms last less than 1 hour while, 81.6$\%$ and 90.8$\%$ last less than 3 and 8 hours respectively \textbf{(Fig.~\ref{fig:21xx})}. Similarly, 59$\%$ of the differential RTT alarms last less than 1 hour while 81.$\%$ and 90.3$\%$ last less than 3 and 6 hours respectively.
%We calculate the duration of an alarm from the first report of the anomaly detection module, until either the anomaly stops or, the reference converges near to the current observed values.
\deleted[id=Alex]{Long lasted alarms are an indication either of permanent routing changes affecting the link or of a strong alarm corrupting the reference value, like in \textbf{Fig.~\ref{fig:36}}. Under both cases the alarm will continue to be reported until the reference converges to new observations.}

\added[id=Alex]{The cause of network disruptions in colos is usually short lived yet traffic patterns can be affected for multiple hours. For instance, for the AMS-IX outage described in \textbf{Section \ref{amsix:outage}}, a 10 minute disruption in the IXP affected the forwarding patterns of colos for 2 hours. Longer lasting alarms are indications either of permanent routing changes (\textbf{Fig.~\ref{fig:43}}) or of a strong alarm corrupting the reference value (\textbf{Fig.~\ref{fig:36}}). In both cases, the alarm will continue to be reported until the reference converges to new observations.}

% It is important to note that during a strong routing alarm, the pattern change of traceroutes passing through the facility links can lead to a change in the median differential RTT pattern. Thus, a false link delay anomaly might be inferred. The delay change and forwarding detection modules complement each other to detect anomalies. Otherwise, some anomalies would not be possible to be inferred applying only one module independently. Without ground truth, it is hard to infer if the differential RTT anomaly was caused by a congested router. 
\replaced[id=Romain]{Our system is sensitive to small pattern changes, but, as 
    described below, we can focus only on significant 
anomalies by ranking alarms based on their deviation from the reference.}
{Since, plenty of alarms are reported by our system \deleted[id=Alex]{in the next subsections} we distinguish strong alarms and rank them based on their deviation from the reference.
}
%\begin{figure}[!h]%
%    \centering
%    \includegraphics[width=0.40\textwidth]{figures/RouteAndRTTAlDurationCDF.png}%
%    \caption{A CDF of the duration of the alarms reported by the anomaly modules.}
%    \label{fig:34}
%\end{figure}

\subsubsection{Ranking Delay Anomalies}

When we detect a differential RTT anomaly we calculate the deviation between the reference and the observed confidence intervals (eq 6 in \textbf{\cite{ref:10}}).
We use this metric to rank the differential RTT anomalies \replaced[id=Alex]{after we remove those where both a delay change and a forwarding anomaly occurred.}
{First, we remove all the anomalies where both a delay change and a forwarding anomaly occurred.}
This is because a change in the forwarding pattern is likely to affect the median RTT and thus to create a false alarm.
%Without ground truth validation it is hard to identify congested routers both dropping and delaying packets. 
\deleted[id=Romain]{By ignoring anomalies where both a forwarding and a delay change occurred only clean delay data remain with deviations that can be ranked. }
Among the top-80 alarms in \textbf{Fig.~\ref{fig:35x}} we observe 3 outstanding cases\deleted[id=Romain]{ with otherwise stable differential patterns}. 

%\begin{figure}%
%    \centering
%    \includegraphics[width=0.42\textwidth]{figures/top150-deviationRTT.png}%
%    \caption{ Deviation of the top-150 RTT alarms.}
%    \label{fig:35}
%\end{figure}

\begin{figure}%
    \centering
    \includegraphics[width=.95\columnwidth]{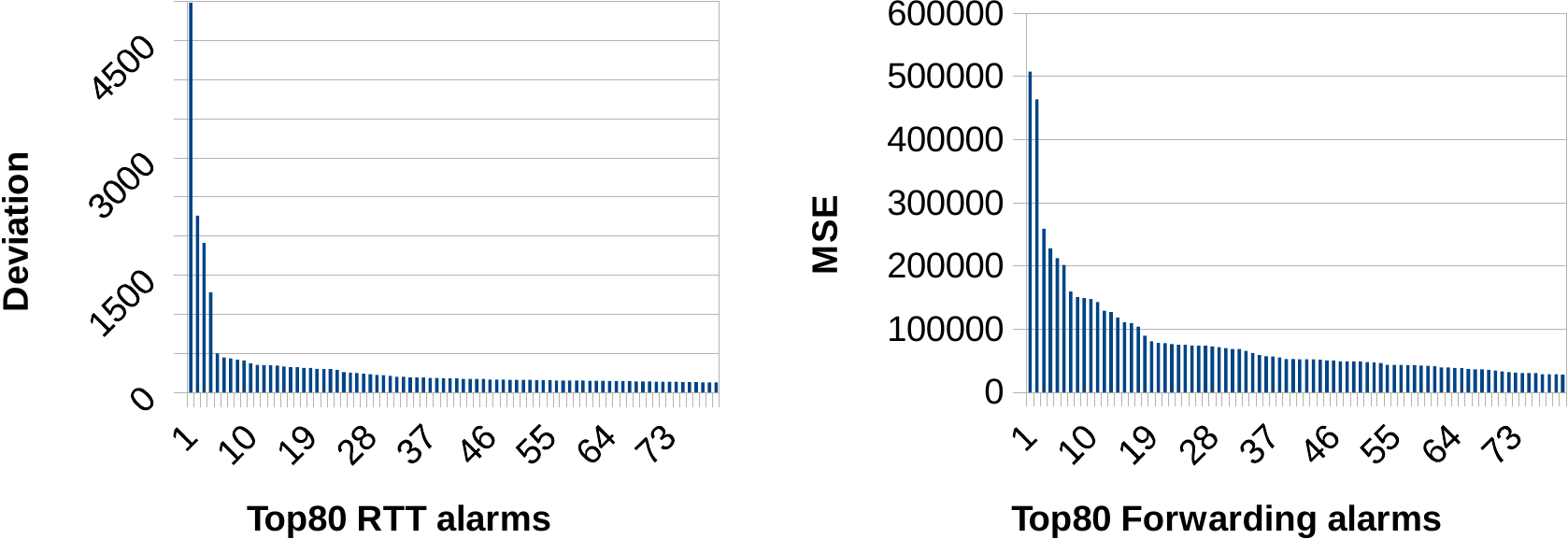}% 0.42
    \caption{Top-80 $RTT_{Diff}$ and forwarding alarms.}
    \label{fig:35x}
\end{figure}

%(Figure~\ref{fig:44} in 
The first case affected \textit{Equinix London LD5} on 2015-9-4 22:00 UTC \deleted[id=Alex]{and lasted}for 2 hours. We observe both intra and inter facility links getting congested \textbf{(Fig.~\ref{fig:36})}. The second was due to a DDoS attack at the end of November (\added[id=Romain]{described in details in }\textbf{Section \ref{section DNS Root Server Attack}}). 
The third stands for \replaced[id=Romain]{delay changes }{an anomaly} on a link that connects \textit{Interxion Frankfurt(FRA1-12)} with \textit{Speedbone Berlin} on 2015-6-8 19:00 UTC.} 
%\todo[nolist, size=\tiny]{No fig here to save space}
%(\textbf{(Figure~\ref{fig:38})}.
Validation from public sources was only possible for the DDoS outage.

\begin{figure}%
    \centering
    \subfloat{\includegraphics[width=0.48\textwidth]{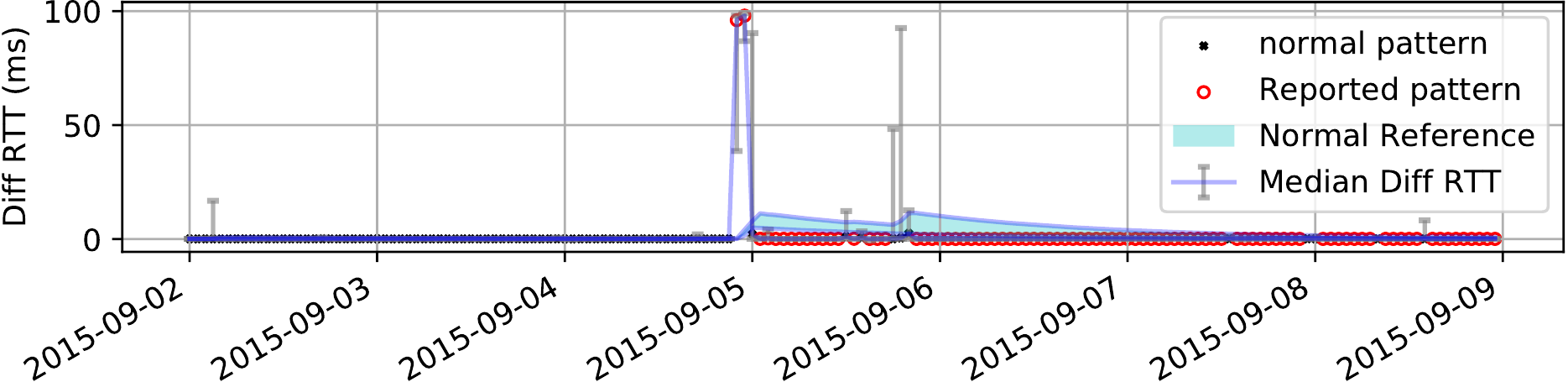} }
    \qquad
    \subfloat{\includegraphics[width=0.48\textwidth]{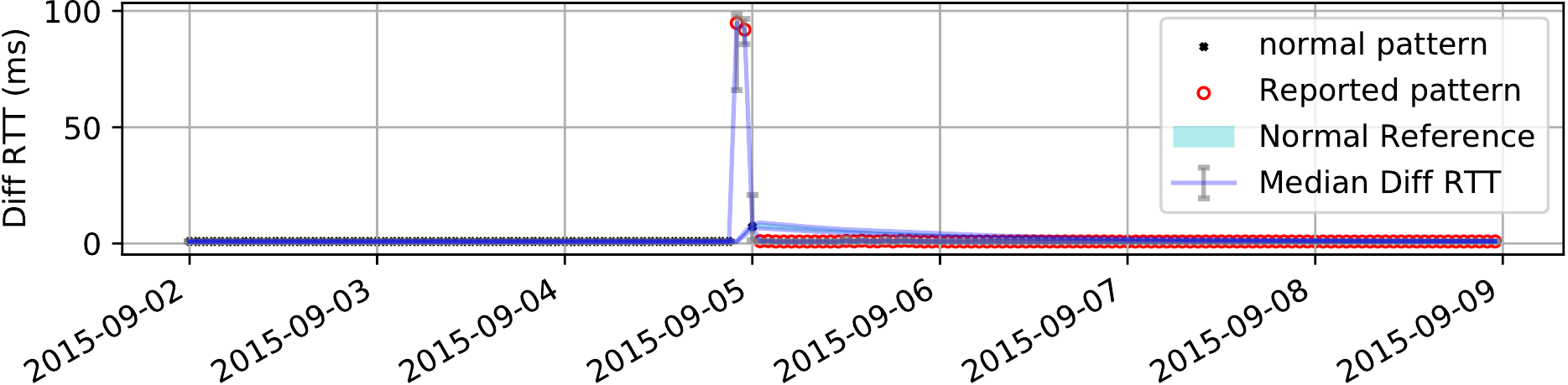} }
    \caption{Top1 delay alarm affecting intra links of \textit{Equinix LD5} (up) and inter links towards \textit{Equinix LD8 (down)}.}
    \label{fig:36}
\end{figure}

%\begin{figure}%
%    \centering
%    \includegraphics[width=0.30\textwidth]{figures/Top1-Fac832To45-RTT.pdf}%
%    \caption{Top1 delay alarm affecting Equinix LD5 links \textit{(fac$\_$id:832)} towards Equinix LD8 \textit{(fac$\_$id:45)}.}
%    \label{fig:37}
%\end{figure}

%
%\begin{figure}[!h]%
%    \centering
%    \subfloat[Forwarding Alarms]{{\includegraphics[width=0.46\textwidth]{figures/Top1-Fac832To45-Route.png}} }%
%    \qquad
%    \subfloat[Differential RTT alarms]{{\includegraphics[width=0.46\textwidth]{figures/Top1-Fac832To45-RTT.png}} }%
%    \caption{Top-1 alarm (b) affecting affecting inter links of Equinix London Slough LD5 (fac$\_$id: 832) to Equinix London Docklands LD8 (fac$\_$id: 45). No change is observed during the affected period in the forwarding pattern (b) }%
%    \label{fig:37}%
%\end{figure}
%
%\begin{figure}[!h]%
%    \centering
%    \subfloat[Forwarding Alarms]{{\includegraphics[width=0.46\textwidth]{figures/Top3-Fac58To155-Route.png}} }%
%    \qquad
%    \subfloat[Differential RTT alarms]{{\includegraphics[width=0.46\textwidth]{figures/Top3-Fac58To155-RTT.png}} }%
%    \caption{Top-3 alarm (b) affecting affecting inter links of Interxion Frankfurt (FRA1$\_$58) to Speedbone Berlin (fac$\_$id: 155). No change is observed during the affected period in the forwarding pattern (b) }%
%    \label{fig:38}%
%\end{figure}

%\FloatBarrier	% force figures to remain in this section

%\newpage
\subsubsection{Ranking Forwarding Anomalies}

\deleted[id=Alex]{To rank a forwarding anomaly}
%of a facility consideration should be taken for \added{both the intra and the inter peering links.
We use the mean squared error to quantify the change in the forwarding patterns of each near-end facility:
%all links connecting it's members, both locally in the same facility and remotely to other colos over the IXP infrastructure. We first tried to rank forwarding anomalies based on the p-value. However, plenty of its values were small, making the classification of strong alarms hard to perform.
%
%MSE(Mean Square Error) \textbf{\cite{ref:59}} is a metric used in statistical modelling to represent the difference between the actual observations and the observation values predicted by the model. 
%In our case, for each facility we define the following MSE:\\
%\added{As a backup plan we use the mean square error (MSE) defined for each facility as}
\[ MSE = \frac{1}{n} \sum_{i=1}^{n} (F_i - \overline{F}_i)^2\]
where $n$ is the number of far-end colos, $F_i$ and $\overline{F}_i$ the observed and the reference forwarding pattern towards far-end colos.

%\begin{figure}%
%    \centering
%    \includegraphics[width=0.42\textwidth]{figures/top150-MSE-Route.png}%
%    \caption{Mean Square Error of the top-150 alarms.}
%    \label{fig:35}
%\end{figure}

\begin{figure}%
\centering
\begin{minipage}{0.42\columnwidth}
   \centering
   \includegraphics[width=\columnwidth]{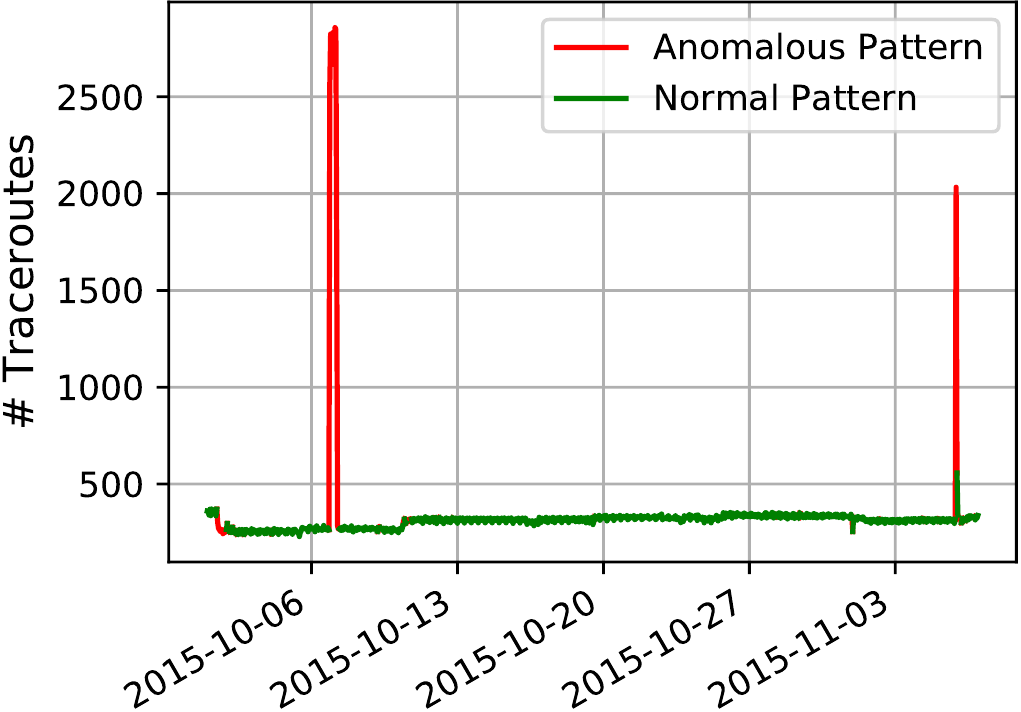}%
   %\captionof{figure}{Top1\&4 forwarding alarms affecting links of \textit{Equinix Frankfurt(FR7)} towards \textit{Equinix Amsterdam(AM7)} at 10-6-22:00 \& 11-5-22:00 UTC.}
   \captionof{figure}{Top1\&4 forwarding alarms from \textit{Equinix Frankfurt(FR7)} to \textit{Equinix Amsterdam(AM7)} at 10-6-22:00 \& 11-5-22:00 UTC.}
   \label{fig:40}
\end{minipage}%
\hfill
\begin{minipage}{0.564\columnwidth}
   \centering
   \includegraphics[width=\columnwidth]{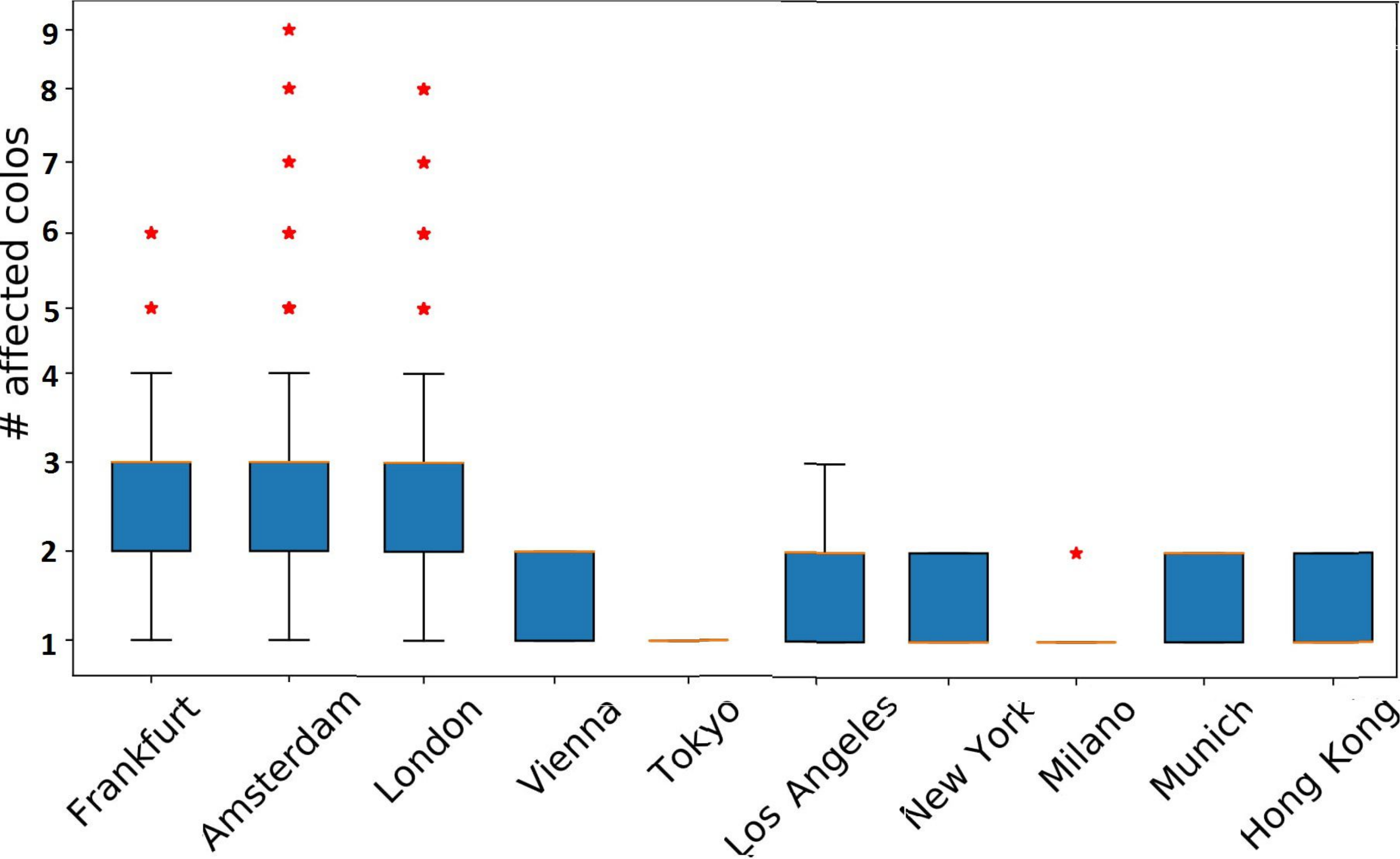}%
   \captionof{figure}{Usual number of affected facilities in the top10 cities. * annotates unusual outages.}
   \label{fig:49}
\end{minipage}
\end{figure}

%From the results of this metric \textbf{(Figure~\ref{fig:39})}, we can clearly observe some outstanding anomalous cases affecting both inter and intra colo links \textbf{(Figure~\ref{fig:40}, \ref{fig:41} and \ref{fig:42} )}.
All top alarms\deleted[id=Alex]{\textbf{(Fig.~\ref{fig:35x} right plot)}} report \replaced[id=Romain]{significant changes in the number of traceroute passing through a link like the example illustrated in }{heavy traffic spikes like those of} \textbf{Fig.~\ref{fig:40}}. 
Such spikes may happen due to \replaced[id=Romain]{inter-facility link failures 
or routing changes.}{BGP policy changes.}
%\todo[nolist, size=\tiny]{Careful! Could it be smth else?}
\replaced[id=Romain]{We hypothesize that the two one-month apart alarms of \textbf{Fig.~\ref{fig:40}}
    are due to maintenance work.}{
The 1 month difference between the two alarms in \textbf{Fig.~\ref{fig:40}} might be an indication of a planned maintenance. }
%When an outage affects a colocation facility we expect to observe traffic loss in some links and increase towards back-up nearby facility crossing links. Such variations are not always visible from our system, since not all facility links have been identified from the facility detection algorithm. For example, we might observe the traffic increase without the traffic loss. This may justify weird traffic bursts in our Figures.

%\begin{figure}[!h]%
%    \centering
%    \includegraphics[width=0.60\textwidth]{figures/Fac214-214-Top2.png}%
%    \caption{Top-2 forwarding alarm affecting intra links of "CE Colo Prague" (fac$\_$id: 214) at 2015-5-31 01:00 UTC.}
%    \label{fig:41}
%\end{figure}
%
%\begin{figure}[!h]%
%    \centering
%    \includegraphics[width=0.60\textwidth]{figures/Fac66-66-Top3.png}%
%    \caption{Top-3 forwarding alarm(left) affecting intra links of "Vienna University Computer Center (Wien)" (fac$\_$d: 66) at 2015-10-8 17:00 UTC.}
%    \label{fig:42}
%\end{figure}

%\FloatBarrier	% force figures to remain in this section

%\newpage
\subsubsection{Detecting Metropolitan Outages}

To further validate our system, we quantify the impact of detected anomalies on colos of the same metropolitan area. 
We first extract each facility's address and use the Google Maps API to obtain the local city's GPS coordinates.
%Using the Vincenty distance module \textbf{\cite{ref:57}}
Then, using those coordinates we calculate the Vincenty distance between each colo and map the ones closer than 50km to the same metropolitan area.
\deleted[id=Romain]{\added[id=Alex]{We ignore periods of time we know there is a problem with the atlas controller or our data collection}}

%Although, we tried to geolocate to street coordinates it was not always feasible.     
%we quantify the impact of alarms on neighboring colos in the same metropolitan area. For this reason, we first extract from PeeringDB the address code of each facility. Using the Google Maps API \textbf{\cite{ref:98}} we translate each address to the GPS coordinates of it's local city. Although, we tried further geolocation to street coordinates, it was not possible for every street address. Afterwards, we used python's implementation of the Vincenty distance \textbf{\cite{ref:57}} to calculate the geographic distance between each facility. Facilities closer than 50 kms were mapped to the same metropolitan area. 

\deleted[id=Alex]{Atlas probe controller failures can cause an incorrect metropolitan alarm.
}
%From our initial findings, we observe a few very strong alarms affecting up to 10 facilities in the same metro area. Since, there is no outage report available online, such alarms are likely to be caused due to RIPE Atlas failures (e.g. probe's controller going down). 
\deleted[id=Alex]{We filter out such cases using the responsibility metric ($r_i$). Our expectation is that during a huge anomaly, the links closer to the source will lose traffic while, other nearby ones will see an increasing load from the back-up paths between colos. We consider observing different links with unusual high and low number of packets, ($r_i > 0.4$ and $r_i < -0.4$) an alternative approach to mitigate database failures. We filter out all other forwarding metropolitan alarms where $r_i < |0.4|$.
}

\added[id=Alex]{We focus only on the metropolitan alarms with the largest impact by filtering out all those that don't include forwarding anomalies with $|r_i|\geq 0.4$}.
\textbf{Table~\ref{table: cities}} annotates the top-10 metropolitan areas based on the number of such observed alarms. 
%We observe cities of North-western Europe as our main alarm contributors. 
Usually most alarms affect only a few facilities in those cities. We \replaced[id=Romain]{found a few instances}{observe though a few hours} where the alarms \replaced[id=Romain]{ spanned across multiple facilities}{were more severe } \textbf{(Fig.~\ref{fig:49})}. As an example, the AMS-IX outage \textbf{\cite{ref:54}} caused forwarding anomalies to links between 8 local facilities\added[id=Romain]{ (see \textbf{Sec.\ref{amsix:outage}}).} 

%\begin{figure}[!h]%
%    \centering
%    \includegraphics[width=0.40\textwidth]{figures/MetroBarPlot.png}%
%    \caption{Top-10 cities based on the number of alarms observed.}
%    \label{fig:48}
%\end{figure}

\begin{table}[]
\centering
\caption{Top10 cities based on the alarms observed.}
\begin{tabular}{|l|l|l|l|l|l|}
\hline
\textit{Cities} & \textit{Frankfurt}   & \textit{Amsterdam} & \textit{London} & \textit{Vienna} & \textit{Tokyo}     \\ \hline
\textit{Alarms} & \textit{459}         & \textit{399}       & \textit{216}    & \textit{125}    & \textit{77}        \\ \hline
\textit{Cities} & \textit{Los Angeles} & \textit{New York}  & \textit{Milano} & \textit{Munich} & \textit{Hong Kong} \\ \hline
\textit{Alarms} & \textit{65}          & \textit{59}        & \textit{57}     & \textit{53}     & \textit{52}        \\ \hline
\end{tabular}
\label{table: cities}
\end{table}

%\begin{figure}[!h]%
%    \centering
%    \includegraphics[width=0.28\textwidth]{figures/Alarm_box_plot.pdf}%
%    \caption{Usual number of affected facilities in the top10 cities. * annotates unusual outages.}
%    \label{fig:49}
%\end{figure}

%Potential real outages due to scheduled maintenances cannot be detected from our system. To identify such cases,we need  information from the participating members, the facility or the IXP operator. On the other hand, by ranking metropolitan areas based on the total number of observed alarms (both filtered and unfiltered) we observe hours around midnight (local time) where there is an increased probability of an alarm. To our surprise, in all areas we observe an increased number of alarms triggering at 9:00 and 17:00. 

% \begin{figure}[!h]%
%     \centering
%     \subfloat[Frankfurt]{{\includegraphics[width=0.46\textwidth]{figures/Frankfurt.png}} }%
%     \qquad
%     \subfloat[Amsterdam]{{\includegraphics[width=0.46\textwidth]{figures/Amsterdam.png}} }%
%     \qquad
%     \subfloat[London]{{\includegraphics[width=0.46\textwidth]{figures/London.png}} }%
%     \qquad
%     \subfloat[Munich]{{\includegraphics[width=0.46\textwidth]{figures/Munich.png}} }%
%     \caption{Hours of the day that each new forwarding anomaly begun. Cities involved: Frankfurt, Amsterdam, London and Munich. Scan period May to December 2015 }%
%     \label{fig:50}%
% \end{figure}

%\FloatBarrier	% force figures to remain in this section

%\newpage
\subsection{\replaced[id=Romain]{Case Studies}{Incidents of Network Outages}}

To validate our system we looked for \replaced[id=Romain]{events in 2015 that were publicly disclosed}{publicly available outages events in 2015 reported} either on mailing lists \textbf{\cite{ref:52,ref:53}} or by network operators. Usually the source of the outage and the entities affected are not made publicly available. In some cases, even the exact duration of the event is not reported, however local news websites and reports can provide good hints for an estimation. We discuss three major events that occurred in May and November 2015 that our system reports.

\subsubsection{Amsterdam Exchange Point (AMS-IX) Outage}
\label{amsix:outage}

\begin{figure}%
   \centering
   \subfloat[Global Switch Amsterdam $\rightarrow$\newline Interxion Science Park]{{\includegraphics[width=0.24\textwidth]{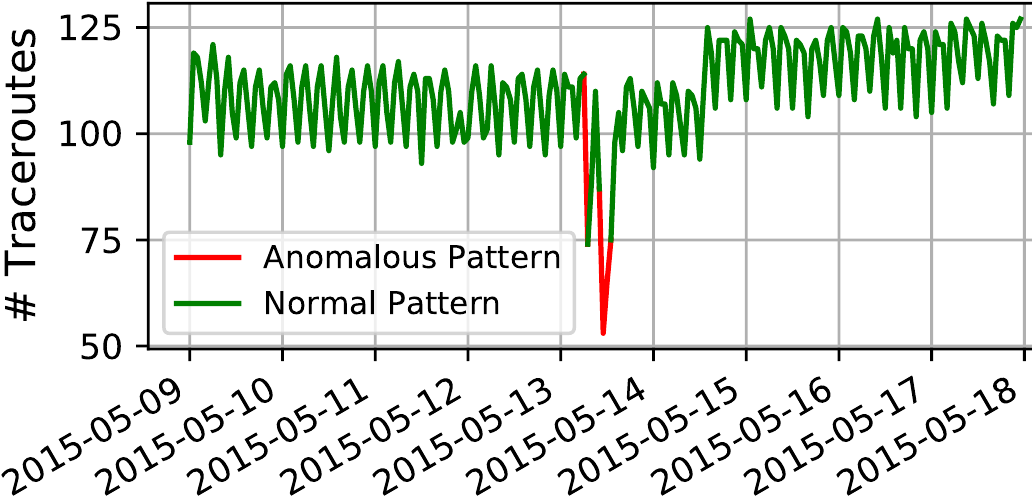}} }%
   \subfloat[Interxion Science Park $\rightarrow$\newline Interxion Frankfurt ]{{\includegraphics[width=0.24\textwidth]{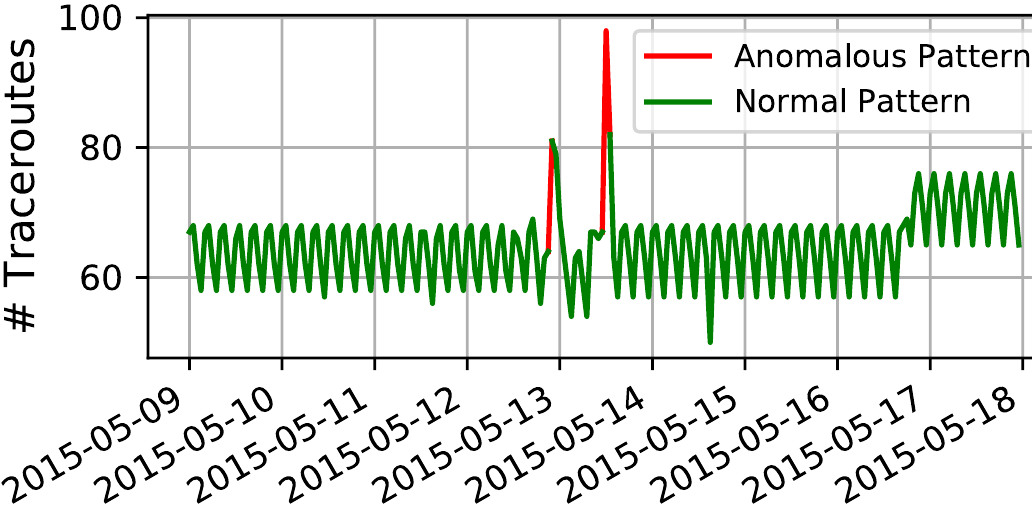}} }%
   \qquad
   \subfloat[Interxion Science Park $\rightarrow$ Equinix AMS South East (AM7) ]{{\includegraphics[width=0.45\textwidth]{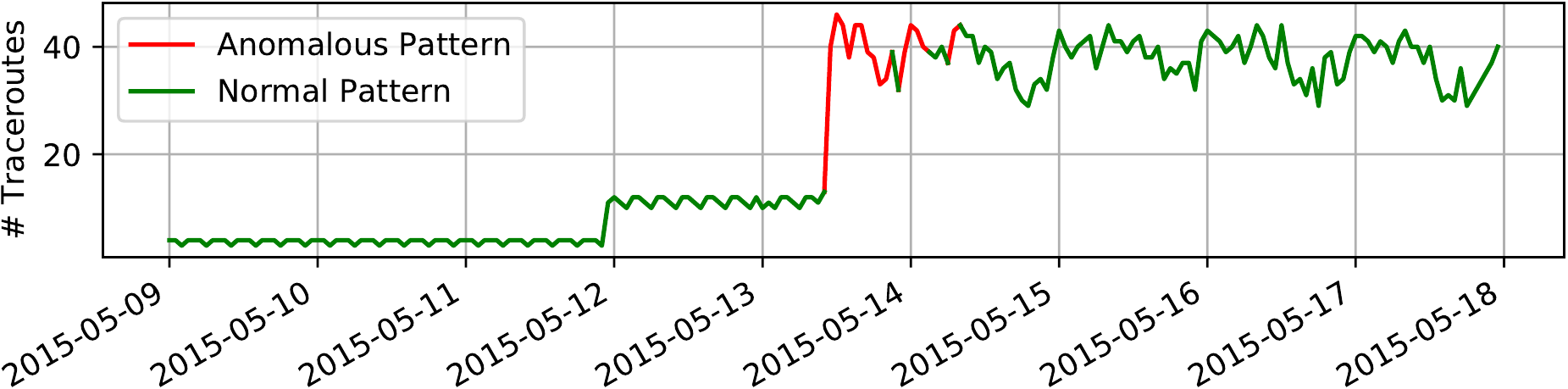}} }%
   \caption{Routing changes observed during the AMS-IX outage.}%
   \label{fig:43}%
\end{figure}

In May-13 between 10:00-12:30 an outage at Amsterdam's core Internet switch infrastructure \textbf{\cite{ref:55}} caused online problems in several parts of the Netherlands. According to news websites the cause was due to a technical fault inside the IXP \textbf{\cite{ref:54}}. 
Using our system we report up to 8 local facilities
\replaced[id=Alex]{with an unusual low and an unsual high forwarding pattern. For example, the}
{having their traffic affected. We observe facility links with an unusual low and an unusual high number of traceroutes %packets
\added[id=Alex]{like those of \textbf{Fig.~\ref{fig:43}}}. The}
outage caused sharp decreases of the number of traceroutes
\replaced[id=Alex]{towards \textit{Interxion Science Park} (\textbf{Fig.~\ref{fig:43} (a)}). Simultaneously, \textit{Science Park} members seem to have used backup paths leading to facilities both inside  (\textbf{Fig.~\ref{fig:43} (c)}) and outside of the country (\textbf{Fig.~\ref{fig:43} (b)}).
We observe that the paths leading inside the country, towards \textit{Equinix AM7}, did not revert back to their usual values probably because of new routes selected after the outage (\textbf{Fig.~\ref{fig:43} (c)}).}
{passing through \textit{Global Switch Amsterdam} inter-links \textbf{(Fig.~\ref{fig:43} (b))} 
and the IXP intra-links in \textit{NIKHEF}.% (\textbf{(Figure~\ref{fig:43}  A and B)}. 
On the other hand, \textit{Interxion Science Park} seems to provide backup paths inside and outside of the country \textbf{(Fig.~\ref{fig:43} (b))}. In addition, we observe that the number of traceroutes for the \textit{Interxion Science Park} did not revert back to its usual values probably because of new routes selected after the outage (\textbf{Fig.~\ref{fig:43} (c)}).}

%\begin{figure}%
   %\centering
   %\subfloat[Global Switch Amsterdam $\rightarrow$\newline Interxion Science Park]{{\includegraphics[width=0.24\textwidth]{AMS_outage/63-104fwd.pdf}} }%
   %\subfloat[Interxion Science Park $\rightarrow$\newline Interxion Frankfurt ]{{\includegraphics[width=0.24\textwidth]{AMS_outage/104-58fwd.pdf}} }%
   %\caption{Routing changes observed during the AMS-IX outage.}%
   %\label{fig:43}%
%\end{figure}

%\begin{figure}[!h]%
%    \centering
%    \subfloat[Global Switch Amsterdam $\rightarrow$ Interxion Science Park]{{\includegraphics[width=0.46\textwidth]{figures/AMS_outage/FAC63To104AMS10-13.png}} }%
%    \qquad
%    \subfloat[NIKHEF $\rightarrow$ Amsterdam NIKHEF Amsterdam]{{\includegraphics[width=0.46\textwidth]{figures/AMS_outage/FAC18To18AMS10-11h.png}} }%
%    \qquad
%    \subfloat[Interxion Science Park $\rightarrow$ Equinix Amsterdam South East]{{\includegraphics[width=0.46\textwidth]{figures/AMS_outage/FAC104To62AMS10-13.png}} }%
%    \qquad
%    \subfloat[Interxion Science Park $\rightarrow$  Interxion Frankfurt ]{{\includegraphics[width=0.46\textwidth]{figures/AMS_outage/FAC104To58AMS10-13.png}} }%
%    \caption{Routing changes observed in inter (A,C,D) and intra links (B) of colocation facilities over the IXP during the AMS-IX outage at 13-May between 10:00-12:30 UTC }%
%    \label{fig:43}%
%\end{figure}

%\FloatBarrier	% force figures to remain in this section
%\newpage

\subsubsection{DDoS attack against DNS Root Servers}
\label{section DNS Root Server Attack}

On November 30, 2015 from 6:50 to 9:30 UTC, and on December 1 from 05:10 to 6:10 UTC, the DNS root servers received an unusually high number of 
\replaced[id=Romain]{spoofed queries \cc{ref:56}.}{\replaced[id=Alex]{well formed queries with spoofed source IP making it unclear from where the traffic had originated \cc{ref:56}. }
{queries. The queries were well formed and the source IP was spoofed making it unclear from where the traffic had originated \mbox{\textbf{\cite{ref:56}}}. 
}}
\replaced[id=Romain]{Each root server has been differently affected by this malicious traffic
\cc{ref:58} but overall the DNS root infrastructure stayed operational during the attack.}{In overall, the anycasted DNS root instances handled well the traffic, although each one had been affected in a different way \cc{ref:58}.}

\begin{figure}%
    \centering
    \subfloat{\includegraphics[width=0.49\textwidth]{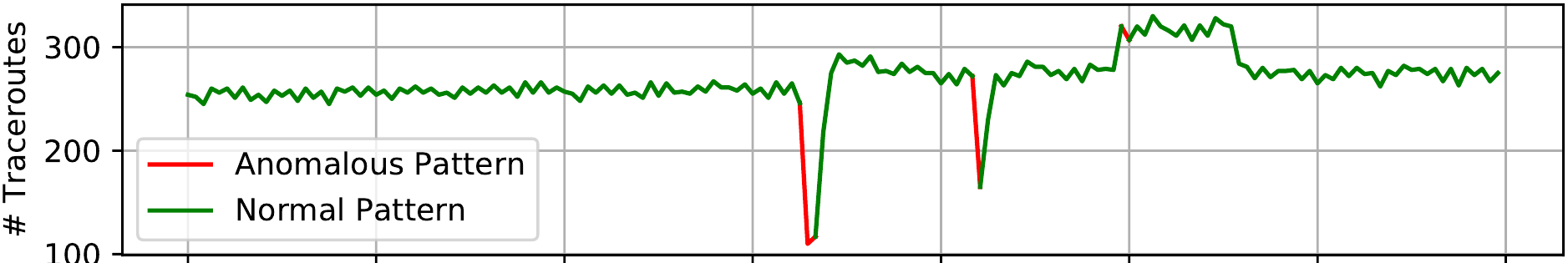} }
    \qquad
    \subfloat{\includegraphics[width=0.49\textwidth]{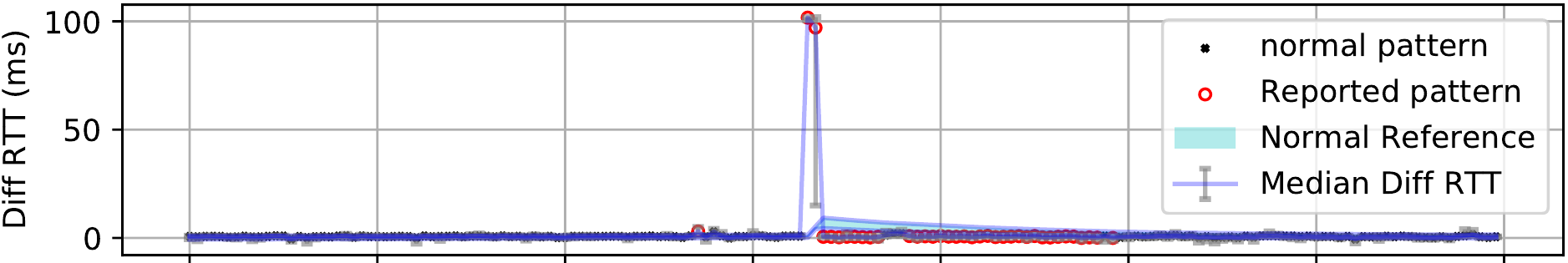} }
    \qquad
    \subfloat{\includegraphics[width=0.49\textwidth]{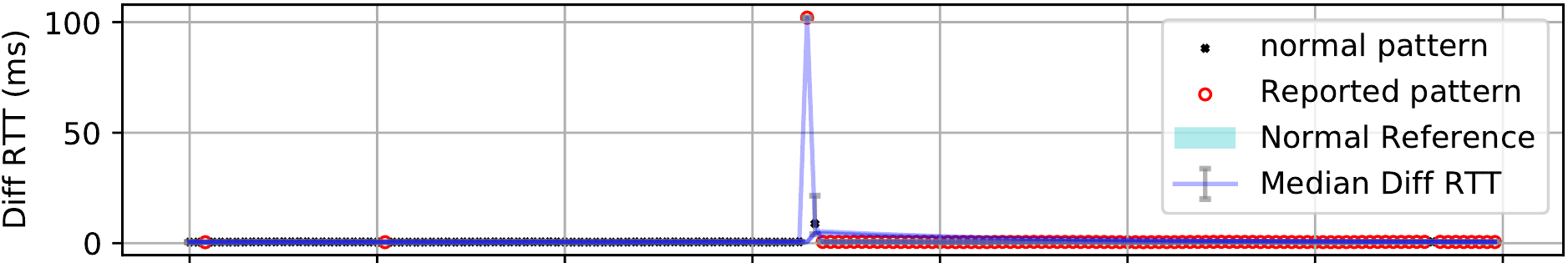} }
    \qquad
    \subfloat{\includegraphics[width=0.49\textwidth]{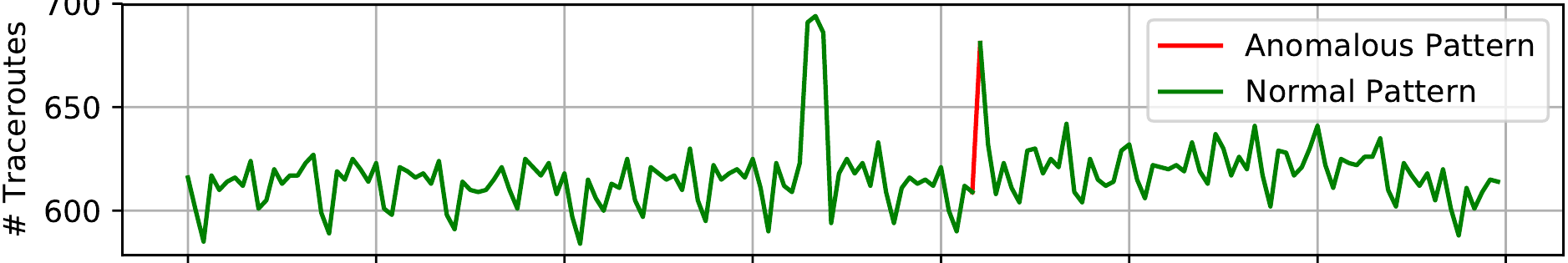} }
    \qquad
    \subfloat{\includegraphics[width=0.49\textwidth]{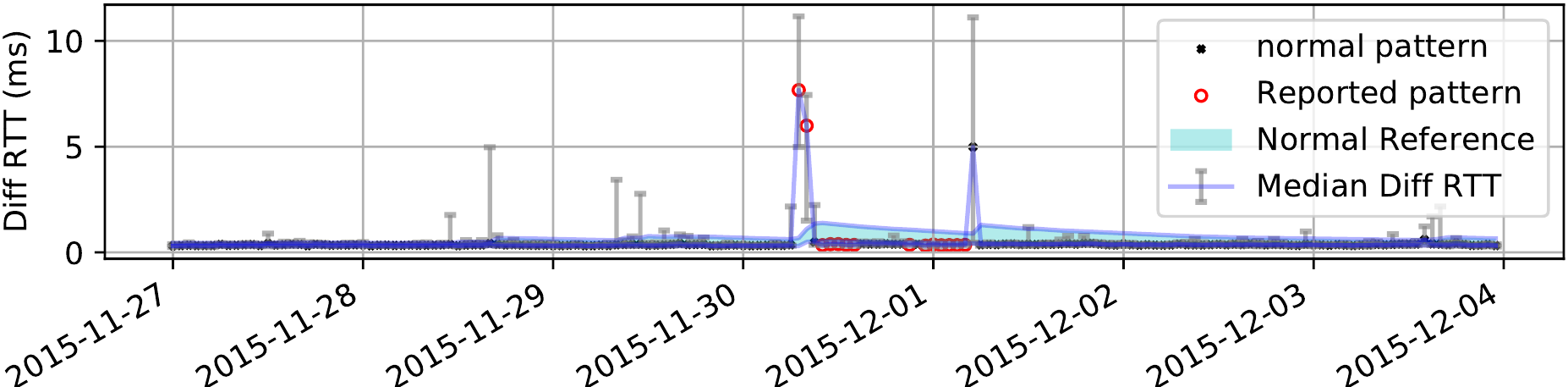} }
    \caption{
    DDoS outage affecting links of \textit{Global switch} towards a)\textit{Digital realty(Wenckebachweg)} and b)\textit{Equinix science Park(AM3)}. Links of c\&d)\textit{Equinix AMS south east(AM7)} towards \textit{Equinix science Park(AM3)} and links of e)\textit{Equinix London Docklands(LD8)} towards \textit{Telehouse Docklands North}.}
    \label{fig:44}
\end{figure}

\replaced[id=Romain]{Our system reports both delay and forwarding anomalies during the attack, mostly for}{We report both delay changes and forwarding anomalies that affected} links in Amsterdam (\textbf{Fig.~\ref{fig:44}A-D}) and London facilities (\textbf{Fig.~\ref{fig:44}E}). \textbf{Fig.~\ref{fig:44}A} indicates that the \textit{Global Switch} links toward \textit{Digital Realty (Wenckebachweg)} Amsterdam got affected during both attacks by a severe forwarding anomaly possibly due to \deleted[id=Alex]{high rates of packet losses from} overloaded routers. At the same time, between 06:00 and 09:00 UTC, we report links of the same facility towards \textit{Equinix Science Park(AM3)} \replaced[id=Alex]{(\textbf{Fig.~\ref{fig:44}B}) with an unusual differential RTT but without an obvious change in the forwarding pattern.}
{ got affected by a change in the differential RTT pattern without the corresponding routing pattern to be affected \textbf{(Fig.~\ref{fig:44}A\&B)}.} 

% \begin{figure}[!h]%
%     \centering
%     \subfloat[Global Switch $\rightarrow$ Equinix Science Park]{{\includegraphics[width=0.46\textwidth]{Ddos_Outage/63-1320RTT.pdf}} }%
%     \qquad
%     \subfloat[Global Switch $\rightarrow$ Equinix Science Park]{{\includegraphics[width=0.46\textwidth]{Ddos_Outage/63-494fwd.pdf}} }%
% %    \qquad
% %    \subfloat[Global Switch $\rightarrow$ Digital Realty]{{\includegraphics[width=0.46\textwidth]{figures/DDoS_Attack/FAC63To494DdosRoute.png}} }%
%     \caption{Routing and differential RTT changes observed in the "Global Switch" facility of Amsterdam. Links towards Digital reality are affected by routing anomalies during both time periods (C) while links towards Science Park are affected only by a differential RTT shift during the first reported period (A,B) }%
%     \label{fig:44}%
% \end{figure}

Links from \textit{Equinix Amsterdam South East(AM7)} towards \textit{Equinix Science Park(AM3)} also experienced high delays, possibly due to the same congested router in the far end facility \textbf{(Fig.~\ref{fig:44}C)}. The increased traffic pattern during the same hour in \textbf{Fig.~\ref{fig:44}D} also \replaced[id=Romain]{confirms}{verifies} that the DNS service \replaced[id=Romain]{hosted near the}{close to} Science Park handled queries of many other unresponsive services. \replaced[id=Romain]{These results corroborate with}{This can be validated by} the results of \textbf{\cite{ref:58}} reporting that many of those services were stressed by sustained traffic during the attack period. It is important also to note that although the two attacks were chronologically close events, the impact of the first one was much stronger for the Amsterdam facilities. During the second attack no RTT change was observed and only routing anomalies were reported.

%\begin{figure}[!h]%
%    \centering
%    \subfloat[Equinix South East $\rightarrow$ Equinix Science Park]{{\includegraphics[width=0.46\textwidth]{figures/DDoS_Attack/FAC62T01320DdosRTT.png}} }%
%    \qquad
%    \subfloat[Global Switch $\rightarrow$ Equinix Science Park]{{\includegraphics[width=0.46\textwidth]{figures/DDoS_Attack/FAC62T01320DdosRoute.png}} }%
%    \caption{Routing and differential RTT changes observed in the "Equinix Amsterdam South East (AM7)" facility of Amsterdam. Links towards "Equinix Amsterdam Science Park (AM3)" are reported to be affected by a differential RTT anomaly during the first Ddos and a forwarding anomaly during the second}%
%    \label{fig:45}%
%\end{figure}

%\begin{figure}[!h]%
%    \centering
%    \subfloat[ Equinix Docklands (LD8) $\rightarrow$ Telehouse Docklands North]{{\includegraphics[width=0.46\textwidth]{figures/DDoS_Attack/FAC42T034DdosRTT.png}} }%
%    \qquad
%    \subfloat[Equinix Docklands (LD8) $\rightarrow$ Telehouse Docklands North]{{\includegraphics[width=0.46\textwidth]{figures/DDoS_Attack/FAC42T034DdosRoute.png}} }%
%    \qquad
%    \subfloat[Telehouse Docklands West $\rightarrow$ Telehouse Docklands North]{{\includegraphics[width=0.60\textwidth]{figures/DDoS_Attack/FAC835T034DdosRoute.png}} }%
%    \caption{ Routing and differential RTT changes observed towards the "Telehouse - London (Docklands North)" facility of London.}%
%    \label{fig:46}%
%\end{figure}

%\FloatBarrier	% force figures to remain in this section
%\newpage

\subsubsection{Telecity Sovereign House outage}

\begin{figure}%
   \centering
   \includegraphics[width=0.48\textwidth]{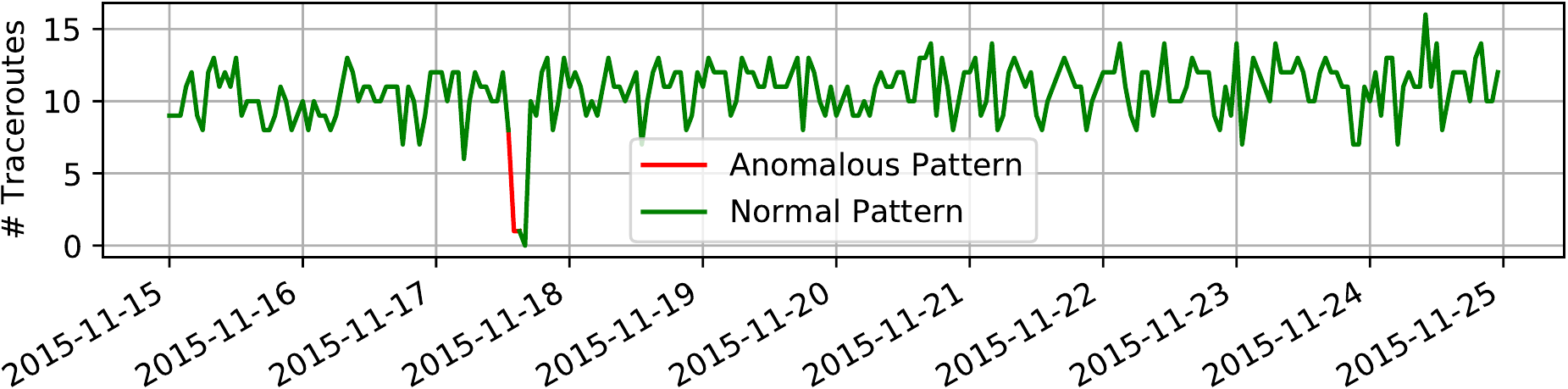}%
   \caption{Outage affecting the links of \textit{Sovereign House} towards \textit{Telehouse-London(Docklands East)} on 2015-11-17 from 14:00 to 15:00 UTC.}
   \label{fig:47}
\end{figure}

On November 17, 2015, a power outage affected London's \textit{Sovereign House} facility where both its primary and secondary supplies failed to start up. No official announcement was made but
%The duration of the outage could not be determined; 
reports from network operators appeared around 2PM local time \textbf{\cite{ref:95}} and continued until the night of the 18th\cc{ref:96}.
Although the visibility of this facility is limited in our datasets, \replaced[id=Romain]{during the outage our system reports a clear drop in the number of traceroutes between London's \textit{Sovereign House} facility and \textit{Telehouse-London(Docklands East)} }{we are able to detect the start of the outage }\deleted[id=Alex]{from links towards ``Telehouse - London (Docklands East)'' that got affected }\textbf{(Fig.~\ref{fig:47})}.
\added[id=Romain]{This illustrates the benefits of the our system
to detect outages although the number of traceroutes crossing the facility might be low.}

Using the proposed colocation facility detection and anomaly detection algorithms, we can evaluate the impact of these events on the physical infrastructure. 
These insights are not available to previous anomaly detection work \textbf{\cite{ref:10}} as they work only at the IP-layer.

%------------------------------------ CHAPTER_7 ------------------------------------------------
\section{Limitations}
\label{chapter: Limitations}

%\replaced[id=Romain]{In this section we discuss the limitations of our proposal.}
%{ In this chapter we discuss the limitations of the facility discovery and the live monitoring.}
%This work aims to detect anomalies at facilities using data plane information. Although we achieved our goal, there are some imperfections in the facility discovery and some limitations in live monitoring.   

\subsection{Facility Identification}
\label{section: Fac Alg Limitations}

In large cities multiple facilities may be connected via cross connect links. When constraining the facility search through alias resolution, the alias ASN may be a member of another facility\replaced[id=Alex]{. For example, in \textbf{Fig.\ref{fig:13}D} \replaced[id=Romain]{if $ASC$ is observed in traceroutes }{if the peering interconnection is established by $ASC$ }then the identification of the near-end \replaced[id=Romain]{may}{will} fail: }
{(e.g., like in \textbf{Fig.~\ref{fig:13}D}). In this case, if the peering interconnection is established by $ASC$ the algorithm will fail: }
$F(ASA) \cap F(ASC) = \{ \O \}$.

\replaced[id=Romain]{
When searching for the far-end facility, we assume that the AS connected to the IXP
is given by the second IP in the far-end IP pair.
Our current implementation maps IP addresses to ASN by simply matching the longest prefix
from Routeviews data.
But if this IP belongs to an inter-AS link we might infer a wrong ASN.
This issue could be addressed by methods like MAP-IT\cc{mapit:imc16} which is
something we are planning to investigate in future work.}
{
MPLS possible used in the IXP border routers may also affect the facility decision. %\textbf{\cite{ref:67}}. 
To recall, when searching the far-end facility the AS connected interface does not appear in the path. As a last resort, we assume that the AS appears in the next hop since the IXP router is an ingress router to it's network. In the case of MPLS though, a different AS may appear instead \cc{vanaubel2017through}.}

%\begin{figure}%
%    \centering
%    \includegraphics[width=0.42\textwidth]{figures/MPLS.png}%
%    \caption{ MPLS problem. IPB is unclear if it belongs to ASB or ASC prefix. }
%    \label{fig:17}
%\end{figure}

\subsection{Live Monitoring Limitations}
\replaced[id=Romain]{
Our system requires multiple vantage points  and 
a common repository to collect all traceroute results (i.e. Atlas probes and controllers
for our current implementation).
Outages appearing close to the data collection may prevent timely access to the data
thus impair the performance of our system.
This case happened during the AMS-IX outage as the RIPE infrastructure is mainly
located in Amsterdam\cc{ref:55}.
}{
\replaced[id=Alex]{In our implementation, we monitored facilities at 1 hour intervals. Although, it is possible to monitor at smaller intervals, RIPE Atlas suffers from limitations when real-time measurements are involved.}
{ Using our system we monitored facilities at 1 hour intervals. RIPE Atlas performs the DNS measurements every 30 minutes. Although, it may be theoretically possible to monitor at smaller intervals RIPE Atlas suffers from some limitations when real-time measurements are involved.}
When an Atlas probe connects to the Internet, a registration server assigns a ``probe-controller'' to the candidate probe \cc{ref:70}. The controller reports the probe's measurements results back to the server, so that they become publicly available. In the case of a controller disruption the probe might have performed the measurements but be unable to make them public, e.g., like in the case of the AMS outage \cc{ref:55}). \added[id=Alex]{Although this was an actual outage,} it's looming of whether the probe is considered connected or disconnected during that time. A live system monitoring crucial destinations may incorrectly infer it as an anomaly while, in reality, this is just a problem between the probe and the controller.}

\section{Conclusions}
\label{chapter:Conclusion and Future_Work}
\replaced[id=Romain]{This work leverages large scale traceroute measurements to
monitor the intricate peering world of colocation facilities.
We devised a system that maps colocation facilities to traceroute data and 
monitors delay and forwarding anomalies at the facility level.
The proposed system enables us to go beyond the usual IP-level monitoring as it 
offers unique inter and intra facility monitoring capabilities.
To demonstrate its benefits we analyzed eight months of data from the RIPE Atlas measurement
platform and reported important outages caused by DoS attacks, power outages, and 
mis-operations. 
We found outages that can span across up to eight facilities and last several hours.
These results provide new insights about the physical locations of facility outages
which are crucial for a better understanding of the peering ecosystem and reliable
connectivity. Furthermore, they provide yet another input for operators to configure their traffic and avoid outages. 
}{In this work, we focus on data plane measurements to shed some light on the  peering world of colocation facilities. Using RIPE Atlas, we first map the colos that traceroutes traverse through and then monitor the inter and intra facility links to detect forwarding and delay anomalies. We scan the Atlas built-in measurements between May and December 2015 where we detect a lot, above 10k of anomaly reports. Ranking the anomalies though, we report only a few that can be considered as actual outages. In the top list of our differential alarms, we observe the DNS attack at the end of November and notice the AMS outage of May affecting the traffic of up to 8 colos in the same area.   }

\deleted[id=Romain]{
In the future, a combination of a data-plane focused and control-plane focused system can yield more complete results from the visibility perspective.
}

\bibliographystyle{IEEEtran} % ref in written order
\bibliography{infocom-bib}

\begin{thebibliography}{10}
\providecommand{\url}[1]{#1}
\csname url@samestyle\endcsname
\providecommand{\newblock}{\relax}
\providecommand{\bibinfo}[2]{#2}
\providecommand{\BIBentrySTDinterwordspacing}{\spaceskip=0pt\relax}
\providecommand{\BIBentryALTinterwordstretchfactor}{4}
\providecommand{\BIBentryALTinterwordspacing}{\spaceskip=\fontdimen2\font plus
\BIBentryALTinterwordstretchfactor\fontdimen3\font minus
  \fontdimen4\font\relax}
\providecommand{\BIBforeignlanguage}[2]{{%
\expandafter\ifx\csname l@#1\endcsname\relax
\typeout{** WARNING: IEEEtran.bst: No hyphenation pattern has been}%
\typeout{** loaded for the language `#1'. Using the pattern for}%
\typeout{** the default language instead.}%
\else
\language=\csname l@#1\endcsname
\fi
#2}}
\providecommand{\BIBdecl}{\relax}
\BIBdecl

\bibitem{ref:15}
C.~Labovitz, S.~Iekel-Johnson, D.~McPherson, J.~Oberheide, and F.~Jahanian,
  ``Internet inter-domain traffic,'' in \emph{ACM SIGCOMM Computer
  Communication Review}, vol.~40, no.~4.\hskip 1em plus 0.5em minus 0.4em\relax
  ACM, 2010, pp. 75--86.

\bibitem{ref:101}
S.~M. Besen and M.~A. Israel, ``The evolution of internet interconnection from
  hierarchy to “mesh”: Implications for government regulation,''
  \emph{Information Economics and Policy}, vol.~25, no.~4, pp. 235--245, 2013.

\bibitem{ref:13}
N.~Chatzis, G.~Smaragdakis, A.~Feldmann, and W.~Willinger, ``There is more to
  ixps than meets the eye,'' \emph{ACM SIGCOMM Computer Communication Review},
  vol.~43, no.~5, pp. 19--28, 2013.

\bibitem{ref:17}
P.~Gill, M.~Arlitt, Z.~Li, and A.~Mahanti, ``The flattening internet topology:
  Natural evolution, unsightly barnacles or contrived collapse?'' in
  \emph{International Conference on Passive and Active Network
  Measurement}.\hskip 1em plus 0.5em minus 0.4em\relax Springer, 2008, pp.
  1--10.

\bibitem{ref:35}
B.~Ager, N.~Chatzis, A.~Feldmann, N.~Sarrar, S.~Uhlig, and W.~Willinger,
  ``Anatomy of a large european ixp,'' \emph{ACM SIGCOMM Computer Communication
  Review}, vol.~42, no.~4, pp. 163--174, 2012.

\bibitem{ref:38}
M.~Z. Ahmad and R.~Guha, ``Studying the effect of internet exchange points on
  internet link delays,'' in \emph{Proceedings of the 2010 Spring Simulation
  Multiconference}.\hskip 1em plus 0.5em minus 0.4em\relax Society for Computer
  Simulation International, 2010, p. 103.

\bibitem{ref:26}
V.~Giotsas, C.~Dietzel, G.~Smaragdakis, A.~Feldmann, A.~Berger, and E.~Aben,
  ``Detecting peering infrastructure outages in the wild,'' in
  \emph{Proceedings of the Conference of the ACM Special Interest Group on Data
  Communication}.\hskip 1em plus 0.5em minus 0.4em\relax ACM, 2017, pp.
  446--459.

\bibitem{ref:16}
``De-cix world record as seen on de-cix news,'' 12 2017.

\bibitem{thousandeyes:decix-outage}
A.~Henthorn-Iwane, ``Understanding internet exchanges via the de-cix outage,''
  \url{https://blog.thousandeyes.com/network-monitoring-de-cix-outage/}.

\bibitem{ref:9}
V.~Giotsas, G.~Smaragdakis, B.~Huffaker, M.~Luckie \emph{et~al.}, ``Mapping
  peering interconnections to a facility,'' in \emph{Proceedings of the 11th
  ACM Conference on Emerging Networking Experiments and Technologies}.\hskip
  1em plus 0.5em minus 0.4em\relax ACM, 2015, p.~37.

\bibitem{cing:infocom03}
K.~G. Anagnostakis, M.~Greenwald, and R.~S. Ryger, ``cing: Measuring
  network-internal delays using only existing infrastructure,'' in
  \emph{INFOCOM 2003}, vol.~3.\hskip 1em plus 0.5em minus 0.4em\relax IEEE,
  2003, pp. 2112--2121.

\bibitem{mahajan:sosp03}
R.~Mahajan, N.~Spring, D.~Wetherall, and T.~Anderson, ``{User-level Internet
  path diagnosis},'' \emph{SOSP'03}, vol.~37, no.~5, pp. 106--119, 2003.

\bibitem{deng:icnp08}
L.~Deng and A.~Kuzmanovic, ``Monitoring persistently congested internet
  links,'' in \emph{Network Protocols, 2008. ICNP 2008. IEEE International
  Conference on}.\hskip 1em plus 0.5em minus 0.4em\relax IEEE, 2008, pp.
  167--176.

\bibitem{ref:10}
R.~Fontugne, C.~Pelsser, E.~Aben, and R.~Bush, ``Pinpointing delay and
  forwarding anomalies using large-scale traceroute measurements,'' in
  \emph{Proceedings of the 2017 Internet Measurement Conference}.\hskip 1em
  plus 0.5em minus 0.4em\relax ACM, 2017, pp. 15--28.

\bibitem{ref:49}
V.~Giotsas, C.~Dietzel, G.~Smaragdakis, A.~Feldmann, A.~Berger, and E.~Aben,
  ``Detecting peering infrastructure outages in the wild,'' in
  \emph{Proceedings of the Conference of the ACM Special Interest Group on Data
  Communication}.\hskip 1em plus 0.5em minus 0.4em\relax ACM, 2017, pp.
  446--459.

\bibitem{ref:41}
R.~Motamedi, B.~Chandrasekaran, B.~Maggs, R.~Rejaie, and W.~Willinger, ``On the
  geography of x-connects,'' Technical Report CIS-TR-2014-02. University of
  Oregon, Tech. Rep., 2014.

\bibitem{ref:27}
I.~Castro, J.~C. Cardona, S.~Gorinsky, and P.~Francois, ``Remote peering: More
  peering without internet flattening,'' in \emph{Proceedings of the 10th ACM
  International on Conference on emerging Networking Experiments and
  Technologies}.\hskip 1em plus 0.5em minus 0.4em\relax ACM, 2014, pp.
  185--198.

\bibitem{ref:45}
R.~Chandra, P.~Traina, and T.~Li, ``Bgp communities attribute,'' Tech. Rep.,
  1996.

\bibitem{ref:2}
``Ripe atlas build-in msms,'' \url{https://atlas.ripe.net/docs/built-in/}.

\bibitem{ref:3}
``Peeringdb,'' \url{https://peeringdb.com/}.

\bibitem{ref:4}
``Caida internet topology data kit,''
  \url{http://www.caida.org/data/internet-topology-data-kit/}.

\bibitem{ref:11}
G.~Nomikos and X.~Dimitropoulos, ``traixroute: Detecting ixps in traceroute
  paths,'' in \emph{International Conference on Passive and Active Network
  Measurement}.\hskip 1em plus 0.5em minus 0.4em\relax Springer, 2016, pp.
  346--358.

\bibitem{ref:5}
``Caida's midar alias resolution,''
  \url{http://www.caida.org/tools/measurement/midar/}.

\bibitem{ref:6}
``Caida iffinder tool,''
  \url{http://www.caida.org/tools/measurement/iffinder/}.

\bibitem{ref:77}
F.~Golkar, T.~Dreibholz, and A.~Kvalbein, ``Measuring and comparing internet
  path stability in ipv4 and ipv6,'' in \emph{Network of the Future (NOF), 2014
  International Conference and Workshop on the}.\hskip 1em plus 0.5em minus
  0.4em\relax IEEE, 2014, pp. 1--5.

\bibitem{ref:78}
G.~Huston, ``ipv6 protocol performance,''
  \url{https://labs.ripe.net/Members/gih/examining-ipv6-performance}.

\bibitem{ref:31}
A.~Dhamdhere, M.~Luckie, B.~Huffaker, A.~Elmokashfi, E.~Aben \emph{et~al.},
  ``Measuring the deployment of ipv6: topology, routing and performance,'' in
  \emph{Proceedings of the 2012 Internet Measurement Conference}.\hskip 1em
  plus 0.5em minus 0.4em\relax ACM, 2012, pp. 537--550.

\bibitem{ref:86}
R.~Steenbergen, ``A practical guide to (correctly) troubleshooting with
  traceroute,'' \emph{North American Network Operators Group}, pp. 1--49, 2009.

\bibitem{ref:80}
B.~Augustin, X.~Cuvellier, B.~Orgogozo, F.~Viger, T.~Friedman, M.~Latapy,
  C.~Magnien, and R.~Teixeira, ``Avoiding traceroute anomalies with paris
  traceroute,'' in \emph{Proceedings of the 6th ACM SIGCOMM conference on
  Internet measurement}.\hskip 1em plus 0.5em minus 0.4em\relax ACM, 2006, pp.
  153--158.

\bibitem{vries:aims15}
W.~de~{Vries}, J.~J. {Santanna}, A.~{Sperotto}, and A.~{Pras}, ``How asymmetric
  is the internet? a study to support the use of traceroute,'' in
  \emph{Intelligent mechanisms for network configuration and security}, ser.
  LNCS, vol. 9122.\hskip 1em plus 0.5em minus 0.4em\relax Springer, June 2015,
  pp. 113--125.

\bibitem{schwartz:infocom10}
Y.~Schwartz, Y.~Shavitt, and U.~Weinsberg, ``{On the diversity, stability and
  symmetry of end-to-end Internet routes},'' in \emph{INFOCOM IEEE Conference
  on Computer Communications Workshops, 2010}.\hskip 1em plus 0.5em minus
  0.4em\relax IEEE, 2010, pp. 1--6.

\bibitem{ref:44}
V.~Kotronis, R.~Kl{\"o}ti, M.~Rost, P.~Georgopoulos, B.~Ager, S.~Schmid, and
  X.~Dimitropoulos, ``Stitching inter-domain paths over ixps,'' in
  \emph{Proceedings of the Symposium on SDN Research}.\hskip 1em plus 0.5em
  minus 0.4em\relax ACM, 2016, p.~17.

\bibitem{ref:54}
``2015-05-13 amsterdam outage,''
  \url{https://nltimes.nl/2015/05/13/outage-amsterdam-internet-hub-affects-much-netherlands}.

\bibitem{ref:52}
``The nanog outage list,''
  \url{https://mailman.nanog.org/mailman/listinfo/nanog}.

\bibitem{ref:53}
``outages list,'' \url{https://puck.nether.net/mailman/listinfo/outages}.

\bibitem{ref:55}
R.~Kisteleki, ``Ams outage as seen from ripe atlas,''
  \url{https://labs.ripe.net/Members/kistel/the-ams-ix-outage-as-seen-with-ripe-atlas}.

\bibitem{ref:56}
R.~S. Operators, ``Events of 2015-11-30,'' 2015.

\bibitem{ref:58}
G.~Moura, R.~Schmidt, W.~B. Heidemann, J.~Vries, M.~Muller, L.~Wei, and
  C.~Hesselman, ``Anycast vs. ddos: Evaluating the november 2015 root dns
  event,'' in \emph{Proceedings of the 2016 Internet Measurement
  Conference}.\hskip 1em plus 0.5em minus 0.4em\relax ACM, 2016, pp. 255--270.

\bibitem{ref:95}
``First reports of sovereign house outage,''
  \url{https://www.theregister.co.uk/2015/11/17/telecity_voip_outage/}.

\bibitem{ref:96}
``2nd day reports of sovereign house outage,''
  \url{https://www.theregister.co.uk/2015/11/18/telecity_outage_fix_failed/}.

\bibitem{mapit:imc16}
A.~Marder and J.~M. Smith, ``Map-it: Multipass accurate passive inferences from
  traceroute,'' in \emph{Proceedings of the 2016 Internet Measurement
  Conference}.\hskip 1em plus 0.5em minus 0.4em\relax ACM, 2016, pp. 397--411.

\end{thebibliography}

\end{document}